\documentclass[12pt,fleqn]{book} 

\usepackage[top=3cm,bottom=3cm,left=3cm,right=3cm,headsep=10pt,a4paper]{geometry} 

\usepackage{graphicx} 
\graphicspath{{graphics/}} 

\usepackage{lipsum} 

\usepackage{tikz} 

\usepackage[english]{babel} 

\usepackage{enumitem} 
\setlist{nolistsep} 

\usepackage[parfill]{parskip} 

\usepackage{booktabs} 

\usepackage{xcolor} 
\usepackage{xcolor-material} 

\usepackage{makecell}

\colorlet{Primary1}{MaterialBlue} 
\colorlet{HintGreen}{MaterialGreen}
\colorlet{CaveatRed}{MaterialRed}

\usepackage{avant} 
\usepackage{mathptmx} 

\usepackage{microtype} 
\usepackage[utf8]{inputenc} 
\usepackage[T1]{fontenc} 

\usepackage[autostyle]{csquotes}

\usepackage[citestyle=numeric,sorting=nyt,sortcites=true,autopunct=true,babel=hyphen,hyperref=true,abbreviate=false,backref=true,backend=biber]{biblatex}
\defbibheading{bibempty}{}

\usepackage{calc} 
\usepackage{makeidx} 
\makeindex 

\usepackage{titletoc} 

\contentsmargin{0cm} 

\titlecontents{part}[0cm]
{\addvspace{20pt}\centering\large\bfseries}
{}
{}
{}

\titlecontents{chapter}[1.25cm] 
{\addvspace{12pt}\large\sffamily\bfseries} 
{\color{Primary1!60}\contentslabel[\Large\thecontentslabel]{1.25cm}\color{Primary1}} 
{\color{Primary1}}
{\color{Primary1!60}\normalsize\;\titlerule*[.5pc]{.}\;\thecontentspage} 

\titlecontents{section}[1.25cm] 
{\addvspace{3pt}\sffamily\bfseries} 
{\contentslabel[\thecontentslabel]{1.25cm}} 
{}
{\hfill\color{black}\thecontentspage} 
[]

\titlecontents{subsection}[1.25cm] 
{\addvspace{1pt}\sffamily\small} 
{\contentslabel[\thecontentslabel]{1.25cm}} 
{}
{\ \titlerule*[.5pc]{.}\;\thecontentspage} 
[]

\titlecontents{figure}[0em]
{\addvspace{-5pt}\sffamily}
{\thecontentslabel\hspace*{1em}}
{}
{\ \titlerule*[.5pc]{.}\;\thecontentspage}
[]

\titlecontents{table}[0em]
{\addvspace{-5pt}\sffamily}
{\thecontentslabel\hspace*{1em}}
{}
{\ \titlerule*[.5pc]{.}\;\thecontentspage}
[]

\titlecontents{lchapter}[0em] 
{\addvspace{15pt}\large\sffamily\bfseries} 
{\color{Primary1}\contentslabel[\Large\thecontentslabel]{1.25cm}\color{Primary1}} 
{}
{\color{Primary1}\normalsize\sffamily\bfseries\;\titlerule*[.5pc]{.}\;\thecontentspage} 

\titlecontents{lsection}[0em] 
{\sffamily\small} 
{\contentslabel[\thecontentslabel]{1.25cm}} 
{}
{}

\titlecontents{lsubsection}[.5em] 
{\normalfont\footnotesize\sffamily} 
{}
{}
{}

\usepackage{fancyhdr} 

\fancypagestyle{plain}{\fancyhead{}} 

\makeatletter
\renewcommand{\cleardoublepage}{
\clearpage\ifodd\c@page\else
\hbox{}
\vspace*{\fill}
\thispagestyle{empty}
\newpage
\fi}

\usepackage{amsmath,amsfonts,amssymb,amsthm} 

\newtheoremstyle{Primary1numbox}
{0pt}
{0pt}
{\normalfont}
{}
{\small\bf\sffamily\color{Primary1}}
{\;}
{0.25em}
{\small\sffamily\color{Primary1}\thmname{#1}\nobreakspace\thmnumber{\@ifnotempty{#1}{}\@upn{#2}}
\thmnote{\nobreakspace\the\thm@notefont\sffamily\bfseries\color{black}---\nobreakspace#3.}} 

\newtheoremstyle{blacknumex}
{5pt}
{5pt}
{\normalfont}
{} 
{\small\bf\sffamily}
{\;}
{0.25em}
{\small\sffamily{\tiny\ensuremath{\blacksquare}}\nobreakspace\thmname{#1}\nobreakspace\thmnumber{\@ifnotempty{#1}{}\@upn{#2}}
\thmnote{\nobreakspace\the\thm@notefont\sffamily\bfseries---\nobreakspace#3.}}

\newtheoremstyle{blacknumbox} 
{0pt}
{0pt}
{\normalfont}
{}
{\small\bf\sffamily}
{\;}
{0.25em}
{\small\sffamily\thmname{#1}\nobreakspace\thmnumber{\@ifnotempty{#1}{}\@upn{#2}}
\thmnote{\nobreakspace\the\thm@notefont\sffamily\bfseries---\nobreakspace#3.}}

\newtheoremstyle{Primary1num}
{5pt}
{5pt}
{\normalfont}
{}
{\small\bf\sffamily\color{Primary1}}
{\;}
{0.25em}
{\small\sffamily\color{Primary1}\thmname{#1}\nobreakspace\thmnumber{\@ifnotempty{#1}{}\@upn{#2}}
\thmnote{\nobreakspace\the\thm@notefont\sffamily\bfseries\color{black}---\nobreakspace#3.}} 
\makeatother

\newcounter{dummy}
\numberwithin{dummy}{section}
\theoremstyle{Primary1numbox}
\newtheorem{theoremeT}[dummy]{Theorem}

\newtheorem{exerciseT}{Exercise}[chapter]
\theoremstyle{blacknumex}
\newtheorem{exampleT}{Example}[chapter]
\theoremstyle{blacknumbox}

\newtheorem{definitionT}{Definition}[section]
\newtheorem{corollaryT}[dummy]{Corollary}
\theoremstyle{Primary1num}

\RequirePackage[framemethod=default]{mdframed} 

\newmdenv[skipabove=7pt,
skipbelow=7pt,
backgroundcolor=black!5,
linecolor=Primary1,
innerleftmargin=5pt,
innerrightmargin=5pt,
innertopmargin=5pt,
leftmargin=0cm,
rightmargin=0cm,
innerbottommargin=5pt]{tBox}

\newmdenv[skipabove=7pt,
skipbelow=7pt,
rightline=false,
leftline=true,
topline=false,
bottomline=false,
backgroundcolor=Primary1!10,
linecolor=Primary1,
innerleftmargin=5pt,
innerrightmargin=5pt,
innertopmargin=5pt,
innerbottommargin=5pt,
leftmargin=0cm,
rightmargin=0cm,
linewidth=4pt]{eBox}

\newmdenv[skipabove=7pt,
skipbelow=7pt,
rightline=false,
leftline=true,
topline=false,
bottomline=false,
linecolor=Primary1,
innerleftmargin=5pt,
innerrightmargin=5pt,
innertopmargin=0pt,
leftmargin=0cm,
rightmargin=0cm,
linewidth=4pt,
innerbottommargin=0pt]{dBox}

\newmdenv[skipabove=7pt,
skipbelow=7pt,
rightline=false,
leftline=true,
topline=false,
bottomline=false,
linecolor=gray,
backgroundcolor=black!5,
innerleftmargin=5pt,
innerrightmargin=5pt,
innertopmargin=5pt,
leftmargin=0cm,
rightmargin=0cm,
linewidth=4pt,
innerbottommargin=5pt]{cBox}

\newenvironment{example}{\begin{exampleT}}{\hfill{\tiny\ensuremath{\blacksquare}}\end{exampleT}}

{\AtBeginEnvironment{theorem}{\parskip=0pt\parsep=0pt\partopsep=0pt}}
{\AtBeginEnvironment{exercise}{\parskip=0pt\parsep=0pt\partopsep=0pt}}
{\AtBeginEnvironment{definition}{\parskip=0pt\parsep=0pt\partopsep=0pt}}
{\AtBeginEnvironment{example}{\parskip=0pt\parsep=0pt\partopsep=0pt}}
{\AtBeginEnvironment{corollary}{\parskip=0pt\parsep=0pt\partopsep=0pt}}

\newenvironment{remark}{\par\vspace{10pt}\small 
\begin{list}{}{
\leftmargin=35pt 
\rightmargin=25pt}\item\ignorespaces 
\makebox[-2.5pt]{\begin{tikzpicture}[overlay]
\node[draw=Primary1!60,line width=1pt,circle,fill=Primary1!25,font=\sffamily\bfseries,inner sep=2pt,outer sep=0pt] at (-15pt,0pt){\textcolor{Primary1}{R}};\end{tikzpicture}} 
\advance\baselineskip -1pt}{\end{list}\vskip5pt} 

\newenvironment{hint}{\par\vspace{10pt}\small 
\begin{list}{}{
\leftmargin=35pt 
\rightmargin=25pt}\item\ignorespaces 
\makebox[-2.5pt]{\begin{tikzpicture}[overlay]
\node[draw=HintGreen!60,line width=1pt,circle,fill=HintGreen!25,font=\sffamily\bfseries,inner sep=2pt,outer sep=0pt] at (-15pt,0pt){\textcolor{HintGreen}{H}};\end{tikzpicture}} 
\advance\baselineskip -1pt}{\end{list}\vskip5pt} 

\newenvironment{caveat}{\par\vspace{10pt}\small 
\begin{list}{}{
\leftmargin=35pt 
\rightmargin=25pt}\item\ignorespaces 
\makebox[-2.5pt]{\begin{tikzpicture}[overlay]
\node[draw=CaveatRed!60,line width=1pt,circle,fill=CaveatRed!25,font=\sffamily\bfseries,inner sep=2pt,outer sep=0pt] at (-15pt,0pt){\textcolor{CaveatRed}{C}};\end{tikzpicture}} 
\advance\baselineskip -1pt}{\end{list}\vskip5pt} 

\makeatletter
\renewcommand{\@seccntformat}[1]{\llap{\textcolor{Primary1}{\csname the#1\endcsname}\hspace{1em}}}
\renewcommand{\section}{\@startsection{section}{1}{\z@}
{-4ex \@plus -1ex \@minus -.4ex}
{1ex \@plus.2ex }
{\normalfont\large\sffamily\bfseries}}
\renewcommand{\subsection}{\@startsection {subsection}{2}{\z@}
{-3ex \@plus -0.1ex \@minus -.4ex}
{0.5ex \@plus.2ex }
{\normalfont\sffamily\bfseries}}
\renewcommand{\subsubsection}{\@startsection {subsubsection}{3}{\z@}
{-2ex \@plus -0.1ex \@minus -.2ex}
{.2ex \@plus.2ex }
{\normalfont\small\sffamily\bfseries}}
\renewcommand\paragraph{\@startsection{paragraph}{4}{\z@}
{-2ex \@plus-.2ex \@minus .2ex}
{.1ex}
{\normalfont\small\sffamily\bfseries}}

\newcommand{\@mypartnumtocformat}[2]{%
\setlength\fboxsep{0pt}%
\noindent\colorbox{Primary1!20}{\strut\parbox[c][.7cm]{\ecart}{\color{Primary1!70}\Large\sffamily\bfseries\centering#1}}\hskip\esp\colorbox{Primary1!40}{\strut\parbox[c][.7cm]{\linewidth-\ecart-\esp}{\Large\sffamily\centering#2}}}%
\newcommand{\@myparttocformat}[1]{%
\setlength\fboxsep{0pt}%
\noindent\colorbox{Primary1!40}{\strut\parbox[c][.7cm]{\linewidth}{\Large\sffamily\centering#1}}}%
\newlength\esp
\newlength\ecart
\def\@part[#1]#2{%
\ifnum \c@secnumdepth >-2\relax%
\refstepcounter{part}%
\addcontentsline{toc}{part}{\texorpdfstring{\protect\@mypartnumtocformat{\thepart}{#1}}{\partname~\thepart\ ---\ #1}}
\else%
\addcontentsline{toc}{part}{\texorpdfstring{\protect\@myparttocformat{#1}}{#1}}%
\fi%
\startcontents%
\markboth{}{}%
{\thispagestyle{empty}%
\begin{tikzpicture}[remember picture,overlay]%
\node at (current page.north west){\begin{tikzpicture}[remember picture,overlay]%
\fill[Primary1!20](0cm,0cm) rectangle (\paperwidth,-\paperheight);
\node[anchor=north] at (4cm,-3.25cm){\color{Primary1!40}\fontsize{220}{100}\sffamily\bfseries\@Roman\c@part};
\node[anchor=south east] at (\paperwidth-1cm,-\paperheight+1cm){\parbox[t][][t]{8.5cm}{
\printcontents{l}{0}{\setcounter{tocdepth}{1}}%
}};
\node[anchor=north east] at (\paperwidth-1.5cm,-3.25cm){\parbox[t][][t]{15cm}{\strut\raggedleft\color{white}\fontsize{30}{30}\sffamily\bfseries#2}};
\end{tikzpicture}};
\end{tikzpicture}}%
\@endpart}
\def\@spart#1{%
\startcontents%
\phantomsection
{\thispagestyle{empty}%
\begin{tikzpicture}[remember picture,overlay]%
\node at (current page.north west){\begin{tikzpicture}[remember picture,overlay]%
\fill[Primary1!20](0cm,0cm) rectangle (\paperwidth,-\paperheight);
\node[anchor=north east] at (\paperwidth-1.5cm,-3.25cm){\parbox[t][][t]{15cm}{\strut\raggedleft\color{white}\fontsize{30}{30}\sffamily\bfseries#1}};
\end{tikzpicture}};
\end{tikzpicture}}
\addcontentsline{toc}{part}{\texorpdfstring{%
\setlength\fboxsep{0pt}%
\noindent\protect\colorbox{Primary1!40}{\strut\protect\parbox[c][.7cm]{\linewidth}{\Large\sffamily\protect\centering #1\quad\mbox{}}}}{#1}}%
\@endpart}
\def\@endpart{\vfil\newpage
\if@twoside
\if@openright
\null
\thispagestyle{empty}%
\newpage
\fi
\fi
\if@tempswa
\twocolumn
\fi}

\newif\ifusechapterimage
\usechapterimagetrue
\newcommand{\thechapterimage}{}%
\newcommand{\chapterimage}[1]{\ifusechapterimage\renewcommand{\thechapterimage}{#1}\fi}%
\def\@makechapterhead#1{%
{\parindent \z@ \raggedright \normalfont
\ifnum \c@secnumdepth >\m@ne
\if@mainmatter
\begin{tikzpicture}[remember picture,overlay]
\node at (current page.north west)
{\begin{tikzpicture}[remember picture,overlay]
\node[anchor=north west,inner sep=0pt] at (0,0) {\ifusechapterimage\includegraphics[width=\paperwidth]{\thechapterimage}\fi};
\draw[anchor=west] (\Gm@lmargin,-9cm) node [line width=2pt,rounded corners=15pt,draw=Primary1,fill=white,fill opacity=0.75,inner sep=15pt]{\strut\makebox[22cm]{}};
\draw[anchor=west] (\Gm@lmargin+.3cm,-9cm) node {\huge\sffamily\bfseries\color{black}\thechapter. #1\strut};
\end{tikzpicture}};
\end{tikzpicture}
\else
\begin{tikzpicture}[remember picture,overlay]
\node at (current page.north west)
{\begin{tikzpicture}[remember picture,overlay]
\node[anchor=north west,inner sep=0pt] at (0,0) {\ifusechapterimage\includegraphics[width=\paperwidth]{\thechapterimage}\fi};
\draw[anchor=west] (\Gm@lmargin,-9cm) node [line width=2pt,rounded corners=15pt,draw=Primary1,fill=white,fill opacity=0.75,inner sep=15pt]{\strut\makebox[22cm]{}};
\draw[anchor=west] (\Gm@lmargin+.3cm,-9cm) node {\huge\sffamily\bfseries\color{black}#1\strut};
\end{tikzpicture}};
\end{tikzpicture}
\fi\fi\par\vspace*{270\p@}}}

\def\@makeschapterhead#1{%
\begin{tikzpicture}[remember picture,overlay]
\node at (current page.north west)
{\begin{tikzpicture}[remember picture,overlay]
\node[anchor=north west,inner sep=0pt] at (0,0) {\ifusechapterimage\includegraphics[width=\paperwidth]{\thechapterimage}\fi};
\draw[anchor=west] (\Gm@lmargin,-9cm) node [line width=2pt,rounded corners=15pt,draw=Primary1,fill=white,fill opacity=0.75,inner sep=15pt]{\strut\makebox[22cm]{}};
\draw[anchor=west] (\Gm@lmargin+.3cm,-9cm) node {\huge\sffamily\bfseries\color{black}#1\strut};
\end{tikzpicture}};
\end{tikzpicture}
\par\vspace*{270\p@}}
\makeatother

\usepackage{hyperref}
\hypersetup{hidelinks,backref=true,pagebackref=true,hyperindex=true,colorlinks=false,breaklinks=true,urlcolor= Primary1,bookmarks=true,bookmarksopen=false,pdftitle={Title},pdfauthor={Author}}
\usepackage{bookmark}
\bookmarksetup{
open,
numbered,
addtohook={%
\ifnum\bookmarkget{level}=0 
\bookmarksetup{bold}%
\fi
\ifnum\bookmarkget{level}=-1 
\bookmarksetup{color=Primary1,bold}%
\fi
}
}
\usepackage{listings} 

\usepackage{caption}
\DeclareCaptionFormat{default}{\small\color{black}#1#2#3}
\usepackage{ltxtable}

\usepackage{mymix}
\usepackage{myconstants}
\usepackage{mylistings}

\begin{document}
\raggedbottom


\begingroup
\thispagestyle{empty}
\begin{tikzpicture}[remember picture,overlay]
\coordinate [below=12cm] (midpoint) at (current page.north);
\node at (current page.north west)
{\begin{tikzpicture}[remember picture,overlay]
\node[anchor=north west,inner sep=0pt] at (0,0) {\includegraphics[width=\paperwidth]{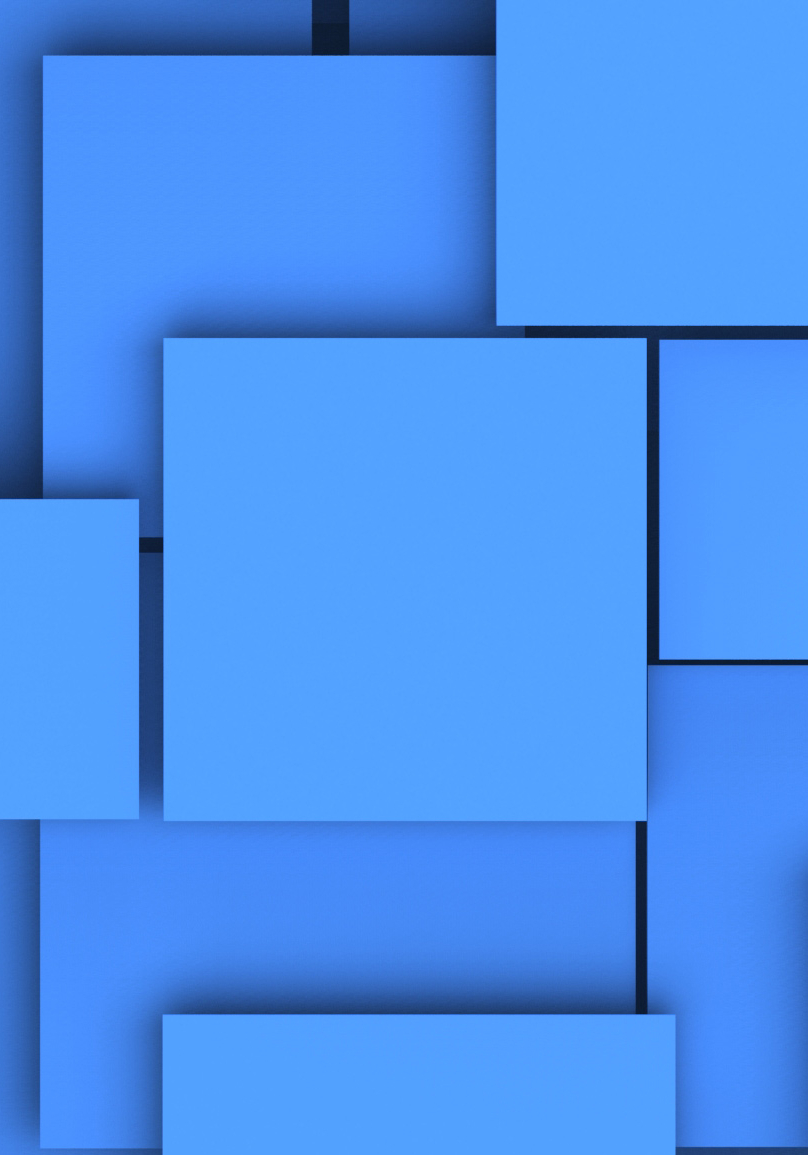}}; 
\draw[anchor=north] (midpoint) node [fill=Primary1!30!white,fill opacity=0.75,text opacity=1,inner sep=1cm]{\Huge\centering\bfseries\sffamily\parbox[c][][t]{\paperwidth}{\centering Introduction to OXPath\\[35pt] 
{\LARGE Ruslan R. Fayzrakhmanov \\ Christopher Michels \\ Mandy Neumann}}}; 
\end{tikzpicture}};xspace
\end{tikzpicture}
\vfill
\endgroup


\newpage
\thispagestyle{empty}
~\vfill
\begin{center}
  {\Huge\bfseries\sffamily Introduction to OXPath} \\
  {\sffamily Version 0.1, June 2018} \\
  \vspace{1cm}
  Ruslan R. Fayzrakhmanov \\ University of Oxford \\ ruslan.fayzrakhmanov@cs.ox.ac.uk \\
  \vspace{1cm}
  Christopher Michels \\ Trier University \\ michelsc@uni-trier.de \\
  \vspace{1cm}
  Mandy Neumann \\ TH Köln \\ mandy.neumann@th-koeln.de
\end{center}
~\vfill

%
%
%
%
\noindent Licensed under the Creative Commons Attribution-NonCommercial-NoDerivatives 4.0 International (CC BY-NC-ND 4.0) License (the ``License''). You may not use this file except in compliance with the License. You may obtain a copy of the License at \url{https://creativecommons.org/licenses/by-nc-nd/4.0/}.
See the License for the specific language governing permissions and limitations under the License.\\ 
%
%
%



\chapterimage{layout/squares_chapter} 

\pagestyle{empty} 

\tableofcontents 

\cleardoublepage 

\pagestyle{fancy} 


\cleardoublepage
\chapter*{Introduction}
\addcontentsline{toc}{chapter}{\textcolor{Primary1}{Introduction}}

A wild growth spurt of the World Wide Web (the Web) flows out of its major integration with different areas of human activities, including business, politics, education, and other essential parts of modern social life.
Thus, the Web has become a versatile medium storing a huge volume of heterogeneous information and data distributed over websites.
In various business and analytical tasks it is considered as a huge source of data long since (e.g., for Competitive Intelligence or price comparison).

Web resources are typically leveraged in ETL (Extract Transform Load) or data wrangling  processes, transforming relevant information identified on Web sources into a structural representation.
\emph{Web data extraction systems} are used in tackling such tasks.
They are applications which automatically and repeatedly extract data from web pages with changing content, delivering the extracted data in a structured form. 
The key component of such systems is a \emph{wrapper} (also known as a ``scraper''), a template, description, or program for extracting relevant data or information, defining the web data extraction strategy.

In fact, Web data extraction is nowadays heavily and proficuously used by various industrial branches.
Electronics retailers, for example, are interested in the daily prices offered by their competitors, so are hotels and supermarket chains.
International construction firms automatically extract tenders from hundreds of websites.
Other sectors have adopted Web data extraction as part of their core business. Among those are flight search engines (such as SkyScanner or Cheapflights) and media intelligence companies (such as Meltwater or Signal).


%
Contemporary web pages with increasingly sophisticated interfaces rival traditional desktop applications for interface complexity and are often called web applications or RIA (Rich Internet Applications).
They often require the execution of JavaScript in a web browser and can call AJAX requests to dynamically generate the content, reacting to user interaction.
From the automatic data acquisition point of view, thus, it is essential to be able to correctly render web pages and mimic user actions to obtain relevant data from the web page content.
Briefly, to obtain data through existing Web interfaces and transform it into structured form, contemporary wrappers should be able to:
1)~interact with sophisticated interfaces of web applications;
2)~precisely acquire relevant data;
3)~scale with the number of crawled web pages or states of web application;
4)~have an embeddable programming API for integration with existing web technologies.
OXPath is a state-of-the-art technology, which is compliant with these requirements and demonstrated its efficiency in comprehensive experiments~\cite{Furche2013-VLDB-OXPath}.
OXPath integrates Firefox for correct rendering of web pages and extends XPath~1.0 for the DOM node selection, interaction, and extraction.
It provides means for converting extracted data into different formats, such as XML, JSON, CSV, and saving data into relational databases.

This tutorial explains main features of the OXPath language and the setup of a suitable working environment.
The guidelines for using OXPath are provided in the form of prototypical examples.

%
%
%

%
%
%


This tutorial is organised as follows:
\begin{itemize}

\item \textbf{Chapter 1, \enquote{Working Environment}},
describes main system requirements for OXPath, different modes in which OXPath wrappers can be executed and its integration into Java applications.

\item \textbf{Chapter 2, \enquote{XPath}}.
It explains main concepts of XPath relevant to understanding the OXPath syntax.

\item \textbf{Chapter 3, \enquote{OXPath}}.
In this chapter, we give a comprehensive description of the OXPath syntax.
A reader will learn different OXPath operators and their use in solving different web data extraction problems.

\item \textbf{Chapter 4, \enquote{OXPath in Action}},
presents different real life use cases mainly by example of bibliographical domain and their Web interfaces.


\item \textbf{Chapter 5, \enquote{Conclusion}},
conclude the material presented in this tutorial.

\end{itemize}

\section*{Conventions}

The examples throughout this tutorial are styled in various ways. These styles differentiate the presented pieces of information. The most important presentation forms illustrated in the following involve listings of OXPath, HTML, and XML source code.




\begin{minipage}{\textwidth}
\begin{lstlisting}[%
    language=oxpath,
    linerange=1-5,
    caption={Example of an OXPath listing}
  ]
doc("https://scholar.google.com")
  //*[@role="search"]//input[@type="text"]/{"OXPath"}
  /../following-sibling::button/{click/}
    //*[@id="gs_ylo_btn"]/{click}
//following::*[@id="gs_ylo_md"]/a[contains(., "2016")]/{click/}
\end{lstlisting}
 \end{minipage}

 \begin{minipage}{\textwidth}
\begin{lstlisting}[%
     language=html,
     caption={Example of an HTML listing}
   ]
<html xmlns="[...]" xml:lang="en">
  <!--[...]-->
    <p>Adaptive Practice of [...]
      <br/>
      <em> Jan Papousek, [...]</em>
      <br/>
      ?!Pages 6-13 [!?
      <a href="uploads/[...].pdf">pdf</a>
      ]
    </p>
  <!--[...]-->
</html>
\end{lstlisting}
 \end{minipage}

 \begin{minipage}{\textwidth}
\begin{lstlisting}[%
     language=xml,
     caption={Example of an XML listing}
   ]
<results>
  <title class="latest">Tim Furche, Georg Gottlob, [...]</title>
  <title>Special Issue: Big Data [...]</title>
  <!--[...]-->
</results>
\end{lstlisting}
 \end{minipage}

Useful caveats, hints and remarks are presented on the go to guide you through this tutorial.


\begin{remark}
Remarks provide additional information and may point to more comprehensive or more advanced content.
\end{remark}

\begin{hint}
Hints contain tips from experience or alternative solutions.
\end{hint}

\begin{caveat}
Caveats warn you of possible misunderstandings, unexpected problems or pitfalls.
\end{caveat}


%
%
%
%

%
%
%


\chapter{Working Environment}

\section{OXPath Client}

A recent version of \OXPath command line interface client, \OXPathCli, (version~\OXPathCliVersion) for \OXPathWithVersion can be downloaded from \OXPathCliUrl.
It is licensed under the 3-Clause BSD License, available at \OXPathLicenseUrl.
\OXPathCli provides a command line interface for executing \OXPath wrappers and saving the extracted data either on the file system in different formats such as XML, CSV, and JSON or in a relational database.
\OXPathWithVersion and \OXPathCliWithVersion can be executed on Linux platforms only, however, other platforms might be supported in future releases.


\section{Requirements}

\OXPath and, therefore, \OXPathCliWithVersion can be only executed on Linux.
It is written in Java and requires the Java Virtual Machine, version~1.7 or higher, to be installed on a Linux machine.
If the execution of \OXPath in X virtual framebuffer (xvfb) is needed, it should be installed, accordingly.
For example, on Debian it can be achieved with the command:
\begin{lstlisting}
sudo apt-get -y install xvfb
\end{lstlisting}


If you used previous versions of OXPath, ensure that the directory \texttt{.diadem/uk/ac/ox/cs/ diadem/webdriver\_env} (in your home directory) is either empty or does not exist at all.
The first execution of \OXPathCli will copy Firefox into this directory.

\section{OXPath Execution}

A list of available parameters for the OXPath client can be generated with one of the following commands:
\begin{lstlisting}
java -jar oxpath-cli.jar
\end{lstlisting}
or
\begin{lstlisting}
java -jar oxpath-cli.jar -h
\end{lstlisting}
A full list of parameters is presented in Table~\ref{tab:oxpath-cli:params}.
\begin{longtable}{l p{3cm} X}
  \caption{\OXPathCli Parameters}\label{tab:oxpath-cli:params} \\
  \toprule\textbf{Parameter} & \textbf{Values} & \textbf{Description} \\ \toprule\endfirsthead
  \toprule\textbf{Parameter} & \textbf{Values} & \textbf{Description} \\ \toprule\endhead
  \texttt{-autocomplete} & & Select first autocomplete suggestion in search forms (disabled by default).\\ \nopagebreak\midrule
  \texttt{-h} & & Print list of parameters.\\ \nopagebreak\midrule
  \texttt{-v} & & Print product version.\\ \nopagebreak\midrule
  -var <file>
  \texttt{-var} & file path & A path to to the file, containing a variable dictionary (Java Properties format) for an oxpath template.\\ \nopagebreak\midrule
  \texttt{-q} & file path & A path to to the file, containing the OXPath expression to evaluate.\\ \nopagebreak\midrule
  \texttt{-o} & file path & The path of the output file in case of XML, CSV or JSON output handlers. If not specified, it outputs results into the default console).\\ \nopagebreak\midrule
  \texttt{-f} & \texttt{xml}, \texttt{json}, \texttt{rscsv}, \texttt{rsjdbc}, \texttt{hcsv}, or \texttt{hjdbc} & Output extracted data in a specified format (\texttt{xml} is default). \texttt{rscsv} and \texttt{rsjdbc} stream basic records into CSV or relational database, respectively. \texttt{hcsv} and \texttt{hjdbc} serialise hierarchy of entities into a single denormalised relation in the form of CSV or into the relational database.\\ \nopagebreak\midrule
  \texttt{-xmlcd} & & If XML mode is enabled, the attribute's value is wrapped into CDATA sections. It is useful if the output contains not allowed XML characters (disabled by default).\\ \nopagebreak\midrule
  \texttt{-jsonarr} & & Use an array for values.\\ \nopagebreak\midrule
  \texttt{-rsent} & string & The name of the entity to be extracted.\\ \nopagebreak\midrule
  \texttt{-rsattrs} & string & List of attributes to be extracted in a form \texttt{a,b,c,d}.\\ \nopagebreak\midrule
  \texttt{-hents} & string & List of attributes to be extracted in a form \texttt{a/b,c/d}\\ \nopagebreak\midrule
  \texttt{-jdbcpar} & file path & XML file with JDBC parameters, such as \texttt{db.driver}, \texttt{db.url}, \texttt{db.user}, \texttt{db.password}, \texttt{db.schema.schema-name} (\texttt{public} is default), \texttt{db.schema.table-name}, \texttt{db.override}, \texttt{db.batch-size}.\\ \nopagebreak\midrule
  \texttt{-mval} & & Allow multiple values per attribute (disabled by default).\\ \nopagebreak\midrule
  \texttt{-ht} & integer & Specify the height of the browser window (default is 800).\\ \nopagebreak\midrule
  \texttt{-wh} & integer & Specify the width of the browser window (default is 1280).\\ \nopagebreak\midrule
  \texttt{-img} & & Enable browser images download (disabled by default).\\ \nopagebreak\midrule
  \texttt{-pl} & & Enable browser's plugins, e.g., Flash and Java (disabled by default).\\ \nopagebreak\midrule
  \texttt{-lat} & float & Latitude (not specified by default).\\ \nopagebreak\midrule
  \texttt{-lon} & float & Longitude (not specified by default).\\ \nopagebreak\midrule
  \texttt{-log} & file path & A file path to the \texttt{log4j} configuration file. If not provided, a default one is applied.\\ \nopagebreak\midrule
  \texttt{-exe} & file path & A file path to the custom executable binary of Firefox (not specified by default).\\ \nopagebreak\midrule
  \texttt{-xvfb} & & Run Firefox in the X virtual framebuffer (disabled by default).\\ \nopagebreak\midrule
  \texttt{-d} & display number & A display port to use in the XVFB mode, e.g., \texttt{:0} (not specified by default).\\ \nopagebreak\bottomrule
\end{longtable}


To execute an OXPath wrapper, along with other parameters the file path to the OXPath wrapper with the parameter \texttt{-q} and an output format using \texttt{-f} need to be specified.
The output format is produced by a specific transformation of the OXPath output, which has a treelike structure and is discussed in Section~\ref{sec:oxpathOutputTree} on page~\pageref{sec:oxpathOutputTree}.

\begin{example}\label{exmpl:dblpGottlob:outputTree}
Given a wrapper presented in Listing~\ref{lst:oxpath:dblpGottlob}, we can extract articles of Georg Gottlob, illustrated in Figure~\ref{fig:dblp-georg}, which includes author names, a title, name of a publication, and pages.
A corresponding OXPath output tree is shown in Figure~\ref{fig:dblp-georg:outTree}.
\begin{lstlisting}[%
language=oxpath,
caption=OXPath wrapper for articles of Georg Gottlob on dblp,
label=lst:oxpath:dblpGottlob
]
doc("http://dblp.dagstuhl.de/pers/hd/g/Gottlob:Georg")
 /.:<collection>
 [
  .:<author="Georg Gottlob">
  :<articles>
  [
   //ul[@class="publ-list"]/li[@class~="entry"]
    /div[@class="data"]:<article>
   [./span[@itemprop="author"]:<author=normalize-space(.)>]
   [./span[@itemprop="name"]:<title=normalize-space(.)>]
   [.//span[@itemprop="isPartOf"][1]
     :<publication=normalize-space(.)>]
   [?./span[@itemprop="pagination"]:<pages=normalize-space(.)>]
  ]
 ]
\end{lstlisting}
\begin{figure}[h]
	\centering
	\includegraphics[width=12cm]{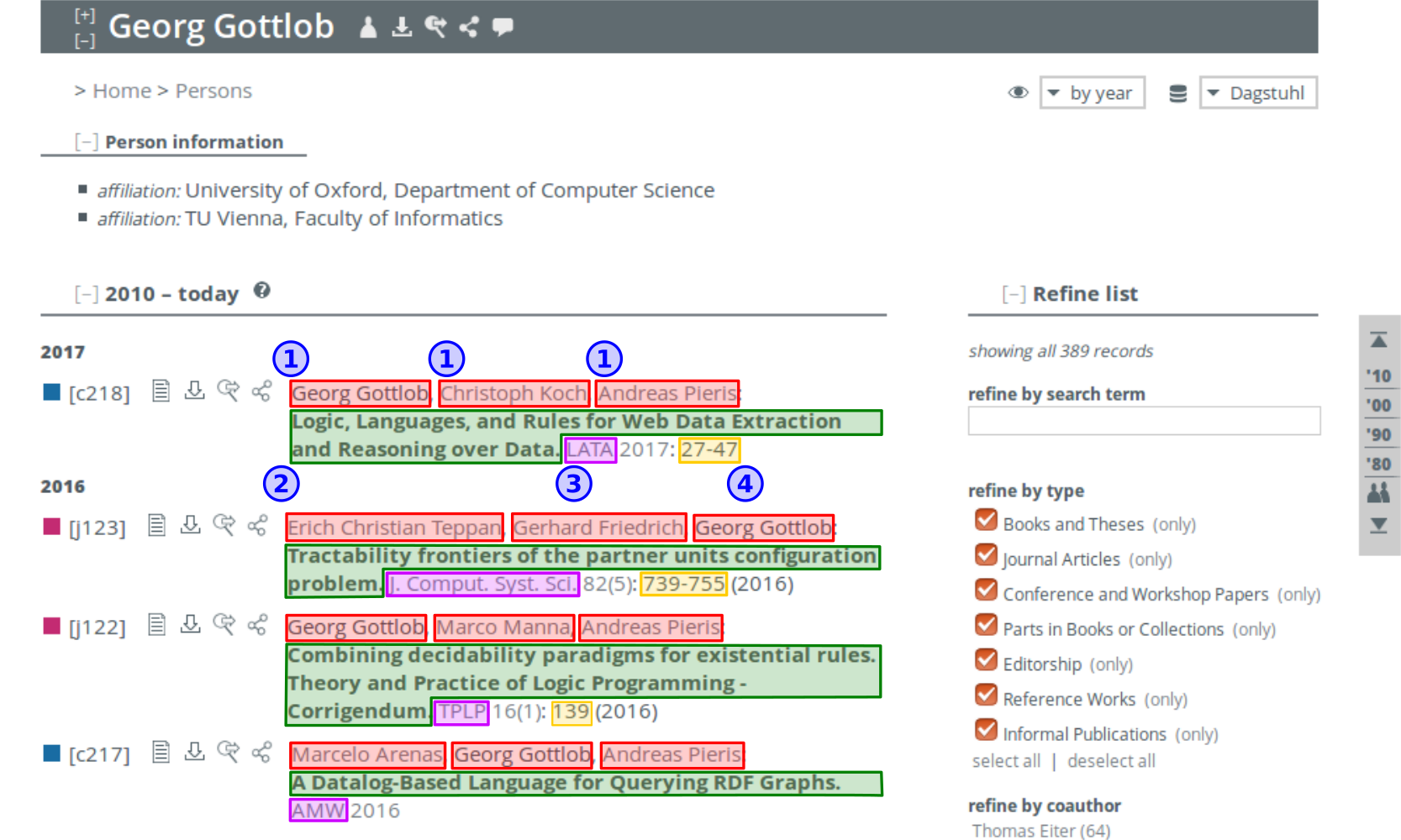}
	\caption{List of papers of Georg Gottlob on dblp.
		Main attributes: (1)~is the author name, (2)~is a title of an article, (3)~is a publication, and (4)~is pages.\label{fig:dblp-georg}}
\end{figure}
\begin{figure}[h]
	\centering
	\includegraphics[width=10cm]{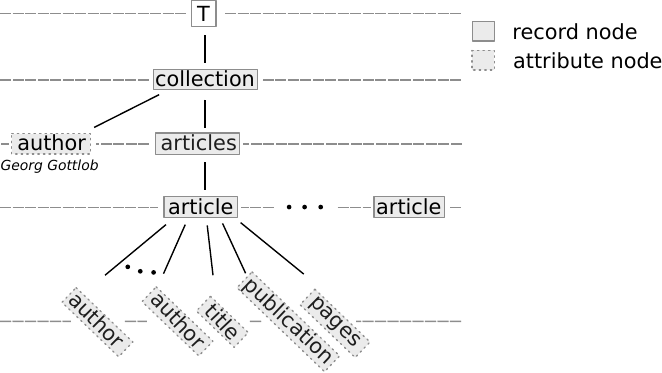}
	\caption{An example OXPath output tree for publications of Georg Gottlob extracted from dblp.\label{fig:dblp-georg:outTree}}
\end{figure}
In this wrapper,
{\color{MaterialRed}\texttt{doc(}}{\color{MaterialBlue}\texttt{"URL"}}{\color{MaterialRed}\texttt{)}}
loads a web page with articles (see Section~\ref{sec:oxpath:rendering} on page~\pageref{sec:oxpath:rendering} about web page rendering).
{\color{MaterialGreen}\texttt{:<name>}}
and
{\color{MaterialGreen}\texttt{:<name=value-exp>}}
are extraction markers (see Section~\ref{sec:extraction-marker} on page~\pageref{sec:extraction-marker}), used for generating the OXPath output tree with record and attribute nodes, respectively.
A DOM node extracted into the output tree is selected with the use of incorporated XPath expressions.
Predicates of the form
\texttt{{\color{MaterialOrange}[\textit{\text{OXPath expression}} ]}}
are used for nesting extracted data nodes;
\texttt{{\color{MaterialOrange}[? \textit{\text{OXPath expression}} ]}}
is an optional predicate denoting an optional attribute to be extracted (see Section~\ref{sec:oxpath:syntsugar-op} on page~\pageref{sec:oxpath:syntsugar-op}).

Thus, this wrapper extracts a tree rooted at $\top$ with a set of articles, in which
{\color{MaterialGreen}\texttt{article/author}} is a multi-valued attribute, and {\color{MaterialGreen}\texttt{pages}} is an optional attribute.
\end{example}

\OXPathCli can transform the OXPath output tree into different formats with the parameter \texttt{-f}.

\subsection{XML Output}\label{sec:xmlOutput}

An XML output can be produced with the use of the parameter \texttt{-f xml} as follows:
\begin{lstlisting}
java -jar oxpath-cli.jar -q <oxpath_expression> -f xml [-mval] [-xmlcd] [-o <output_file>]
\end{lstlisting}
It is a bijective transformation of the output tree into its XML counterpart, in which $\top$ is transformed into the
{\color{MaterialGreen}\texttt{results}} element, all nodes of the original OXPath output tree are XML elements, and each attribute node contains a textual value.

\begin{example}\label{exmpl:dblpGottlob:xml}
The wrapper presented in Listing~\ref{lst:oxpath:dblpGottlob} can be executed with the command of the following form:
\begin{lstlisting}
java -jar oxpath-cli.jar -q <oxpath_expression> -f xml -mval -o <output_file>
\end{lstlisting}
The flag \texttt{-mval} states that we accept same attribute output nodes as siblings in the output tree.
This flag is important in our example as we can have several authors for the same article.
A snippet of the XML output is presented below.
%
\begin{lstlisting}[%
language=xml%,
%caption=A snippet of an example XML output,
%label=lst:xml:dblpGottlob
]
<results>
    <collection>
        <author>Georg Gottlob</author>
        <articles>
            <article>
                <author>Georg Gottlob</author>
                <author>Christoph Koch</author>
                <author>Andreas Pieris</author>
                <title>Logic, Languages, and Rules for Web Data Extraction and Reasoning over Data.</title>
                <publication>LATA</publication>
                <pages>27-47</pages>
            </article>
            ...
        </articles>
    </collection>
</results>
\end{lstlisting}
\end{example}

\subsection{JSON Output}\label{sec:jsonOutput}

A command for producing a JSON output has the following form:
\begin{lstlisting}
java -jar oxpath-cli.jar -q <oxpath_expression> -f json [-mval] [-jsonarr] [-o <output_file>]
\end{lstlisting}
If the parameter \texttt{-mval} is specified, the parameter \texttt{-jsonarr} should also be set.
This enables the serialisation of values of the same attribute and sibling nodes with the same name into a JSON array.

\begin{example}\label{exmpl:dblpGottlob:json} \ 
\begin{lstlisting}
java -jar oxpath-cli.jar -q <oxpath_expression> -f json -mval -jsonarr -o <output_file>
\end{lstlisting}
Using this expression with the OXPath wrapper from Listing~\ref{lst:oxpath:dblpGottlob}, we can generate the following JSON output. 

\begin{lstlisting}[%
language=json%,
%caption=A snippet of an example JSON output,
%label=lst:json:dblpGottlob
]
{
  "collection": [
    {
      "author": ["Georg Gottlob"],
      "articles": [
        {
          "article": [
            {
              "author": [
                "Georg Gottlob",
                "Christoph Koch",
                "Andreas Pieris"],
              "title": ["Logic, Languages, and Rules for Web Data Extraction and Reasoning over Data."],
              "publication": ["LATA"],
              "pages": ["27-47"]
            },
            ...
          ]
        }
      ]
    }
  ]
}
\end{lstlisting}
\end{example}

\subsection{CSV Output}\label{sec:csvOutput}

\OXPathCli supports two CSV transformations.
The first one, \texttt{rscsv}, serialises a specific type of a record node from the OXPath output tree along with its attributes into a set of tuples (i.e., rows in CSV).
This will be discussed further in this section.

The second type of CSV serialisation, \texttt{hcsv}, enables the output of different kinds of record nodes into tuples.
In case of multiple attribute values, those are transformed into a single attribute by joining values into a string with a splitting symbol ``|'';
``\textbackslash'', in turn, is used as an escape symbol.

\paragraph{Simple Record Streaming with rscsv}

For the \texttt{rscsv} transformation the user should specify a node in the OXPath output tree to be identified as a record (a \textit{record node}) and its \textit{attribute nodes} (nodes with string values) by their names.
Each serialised record node is represented in CSV by a list of attributes for each of which this record node is the nearest ancestor in the OXPath output tree.

The \texttt{rscsv} transformation can be invoked with the command as in:
\begin{lstlisting}
java -jar oxpath-cli.jar -q <oxpath_expression> -f rscsv [-mval] [-rsent <record_node_name>] [-rsattrs <attributes>] [-o <output_file>]
\end{lstlisting}
A parameter \texttt{-rsent} is the name of a record node and \texttt{-rsattrs} is a comma-separated list of attributes.

The CSV generation is performed ``on-the-fly'' for nodes extracted by the OXPath processor.
The transformed output is streamed either into a file (if \texttt{-o} is specified) or console (otherwise).
With a low memory footprint this output handler is suitable for processing big streams of data.

\begin{example}\label{exmpl:dblpGottlob:rscsv} \ 
\begin{lstlisting}
java -jar oxpath-cli.jar -q <oxpath_expression> -f rscsv -mval -rsent article -rsattrs author,title,publication,pages -o <output_file>
\end{lstlisting}
With the OXPath wrapper from Listing~\ref{lst:oxpath:dblpGottlob} this \OXPathCli command line serialises attributes 
{\color{MaterialGreen}\texttt{author}},
{\color{MaterialGreen}\texttt{title}},
{\color{MaterialGreen}\texttt{publication}}, and
{\color{MaterialGreen}\texttt{pages}}
-- descendant elements of {\color{MaterialGreen}\texttt{article}} -- from the OXPath output tree and produces the following CSV file.

\begin{lstlisting}[%
language=csv%,
%caption=A snippet of an example CSV output produced by the \texttt{rscsv} transformation,
%label=lst:rscsv:dblpGottlob
]
"id","author","title","publication","pages"
"0","Georg Gottlob|Christoph Koch|Andreas Pieris","Logic, Languages, and Rules for Web Data Extraction and Reasoning over Data.","LATA","27-47"
"1","Erich Christian Teppan|Gerhard Friedrich|Georg Gottlob","Tractability frontiers of the partner units configuration problem.","J. Comput. Syst. Sci.","739-755"
...
\end{lstlisting}
\end{example}

\paragraph{Record Serialisation with hcsv}

In contrast to \texttt{rscsv}, a \texttt{hcsv} transformation serialises a hierarchy of record nodes into a single relation, i.e., a set of tuples with the same schema.
For example, it can serialise a complex entity ``collection'' from the OXPath output tree in Figure~\ref{fig:dblp-georg:outTree} with its attribute ``author'' and containing article records into a single CSV.
Each type of record nodes to be serialised is defined by its path relative to its nearest ancestor record node (a \emph{parent record node}).
A \emph{top record node}, representing a whole entity, is specified with a path relative to the root of the OXPath output tree.
The path is a basic XPath expression of the form \texttt{a/b/.../c}, in which \texttt{a}, \texttt{b}, and \texttt{c} are node names and ``\texttt{/}'' defines a child axis (i.e., parent-child relation between \texttt{a} and \texttt{b} and the same with further elements of the path).
\texttt{hcsv} allows only one path for the top record node, and each record node cannot have several child record nodes with different relative paths.
Following our example, a top record node ``collection'' should be defined by the path \texttt{collection}, while descendant article nodes are with the path \texttt{articles/article}.

Thus, the \texttt{hcsv} mode transforms the hierarchy of record nodes into the first normal form.
Each record node in the serialsed tuples is represented by its sequential ID and attribute nodes, identified in the tree.
A corresponding ID is added to distinguish different entities within the same and in different tuples.
A generated header in the CSV has the following convention for names:
each ID-field has the name of the form \texttt{<record name>\_id}, each attribute name is a ``\texttt{\_}''-separated relative path (see Listing~\ref{lst:hcsv:dblpGottlob}).

A generic command for the \texttt{hcsv} serialisation:
\begin{lstlisting}
java -jar oxpath-cli.jar -q <oxpath_expression> -f hcsv [-mval] -hents <entity_paths> [-o <output_file>]
\end{lstlisting}

The parameter \texttt{-hents} is a comma-separated list of relative paths to record nodes (or the root).
The first path is associated with the main record and each $i+1$-th path is relative to the $i$-th path in the OXPath output tree.
This thus sets a one-to-many relation between $i$-th and $i+1$-th record node.
These are demonstrated in Example~\ref{exmpl:dblpGottlob:hcsv}.

\begin{example}\label{exmpl:dblpGottlob:hcsv}
A snippet of CSV produced by the \texttt{hcsv} output handler for the OXPath expression from Listing~\ref{lst:oxpath:dblpGottlob} and a command line for \OXPathCli are listed below.
\begin{lstlisting}
java -jar oxpath-cli.jar -q <oxpath_expression> -f hcsv -mval -hents collection,articles/article -o <output_file>
\end{lstlisting}

\begin{lstlisting}[%
language=csv%,
%caption=A snippet of an example CSV output produced by the \texttt{hcsv},
%label=lst:hcsv:dblpGottlob
]
"collection_id","collection_author","article_id", "article_publication","article_title","article_pages", "article_author"
"0","Georg Gottlob","0","LATA","Logic, Languages, and Rules for Web Data Extraction and Reasoning over Data.","27-47","Georg Gottlob|Christoph Koch|Andreas Pieris"
"0","Georg Gottlob","1","J. Comput. Syst. Sci.","Tractability frontiers of the partner units configuration problem.","739-755","Erich Christian Teppan|Gerhard Friedrich|Georg Gottlob"
...
\end{lstlisting}
\end{example}

\subsection{Database Output}\label{sec:dbOutput}

As with CSV output (see Section~\ref{sec:csvOutput}), there are two possible database outputs:
1) \texttt{rsjdbc}, the serialisation is similar to \texttt{rscsv},
and 2) \texttt{hjdbc} which is similar to \texttt{hcsv}.

The \texttt{rsjdbc} output can be generated with the following command:
\begin{lstlisting}
java -jar oxpath-cli.jar -q <oxpath_expression> -f rsjdbc [-mval] -rsent <record_node_name> -rsattrs <attributes> -jdbcpar <db_parameters_file>
\end{lstlisting}
A command for the \texttt{hjdbc} transformation has the form:
\begin{lstlisting}
java -jar oxpath-cli.jar -q <oxpath_expression> -f hjdbc [-mval] -hents <entity_paths> -jdbcpar <db_parameters_file>
\end{lstlisting}

A database configuration can be defined with the \OXPathCli parameter \texttt{-jdbcpar}, which specified the file path to an XML file with the structure as in Listing~\ref{lst:oxpath:dbParamsStructure}.

\begin{lstlisting}[%
language=xml,
caption=Structure of the XML file with database parameters,
label=lst:oxpath:dbParamsStructure
]
<diadem>
 <db>
  <driver>...</driver>
  <url>...</url>
  <user>...</user>
  <password>...</password>
  <schema>
   <schema-name>...</schema-name>
   <table-name>...</table-name>
  </schema>
  <override>...</override>
  <batch-size>...</batch-size>
 </db>
</diadem>
\end{lstlisting}

The \texttt{driver} parameter is a fully qualified class name of the JDBC driver for a relational database, e.g., \texttt{org.postgresql.Driver} for Postgres.
\texttt{url}, \texttt{user}, and \texttt{password} specify a JDBC connection to the database.
The \texttt{override} parameter is either \texttt{true} or \texttt{false}.
If it is \texttt{true}, the table will be dropped and created again just before sending the data.
If it is \texttt{false}, the data will be appended to the table. Regardless of the value of \texttt{override}, both schema and table will be always created in case they are not found in the database.
\texttt{batch-size} is an integer value that defines a minimal set of records to be committed at once into the database.

\section{Devising XPath Selectors}

OXPath extends XPath and, therefore, utilises similar syntax and semantics for identifying elements on a web page.
An OXPath user, writing an OXPath expression to extract data from a website, needs to specify all XPath strings both for elements to be interacted with and those to be extracted.
For that the user can use means provided by modern web browsers (e.g., Web Developer Tools in Firefox) as well as additional browsers' plug-ins and add-ons.
Due to the fact that different versions of web browsers can generate differing DOM trees, for writing an XPath expression we recommend to use a browser similar to that integrated into the OXPath engine.
For example, \OXPathWithVersion integrates Firefox~47.0.1\footnote{\url{https://ftp.mozilla.org/pub/firefox/releases/47.0.1/}} with rendering engine Gecko~47.0.

\subsection{Web Developer Tools}

Firefox Developer Tools\footnote{\url{https://developer.mozilla.org/en-US/docs/Tools/Tools_Toolbox}} is a set of means tailored for web developers and give access to various facets of web applications.
Web Console, in particular, can be used for the analysis of the DOM tree (in the Inspector Tab of the toolbox), devising and evaluating XPath expressions (in the Console Tab of the toolbox).
In the Console Tab, the XPath can be evaluated with the command \texttt{\$x()}.

\subsection{Firebug and FirePath}

Firebug\footnote{\url{http://getfirebug.com/}} facilitates the analysis of web applications and, similar to the integrated Firefox Developer Tool, conveys an extensive access to various aspects of web applications executed in the web browser.
FirePath\footnote{\url{https://addons.mozilla.org/En-us/firefox/addon/firepath/}}, an extension for the Firefox add-on Firebug, is a convenient tool to generate and choose an appropriate XPath
for OXPath expressions.
FirePath can be installed with the Add-ons Manager in Firefox.

%

\begin{figure}[t]
\centering
\includegraphics[width=0.9\textwidth]{%
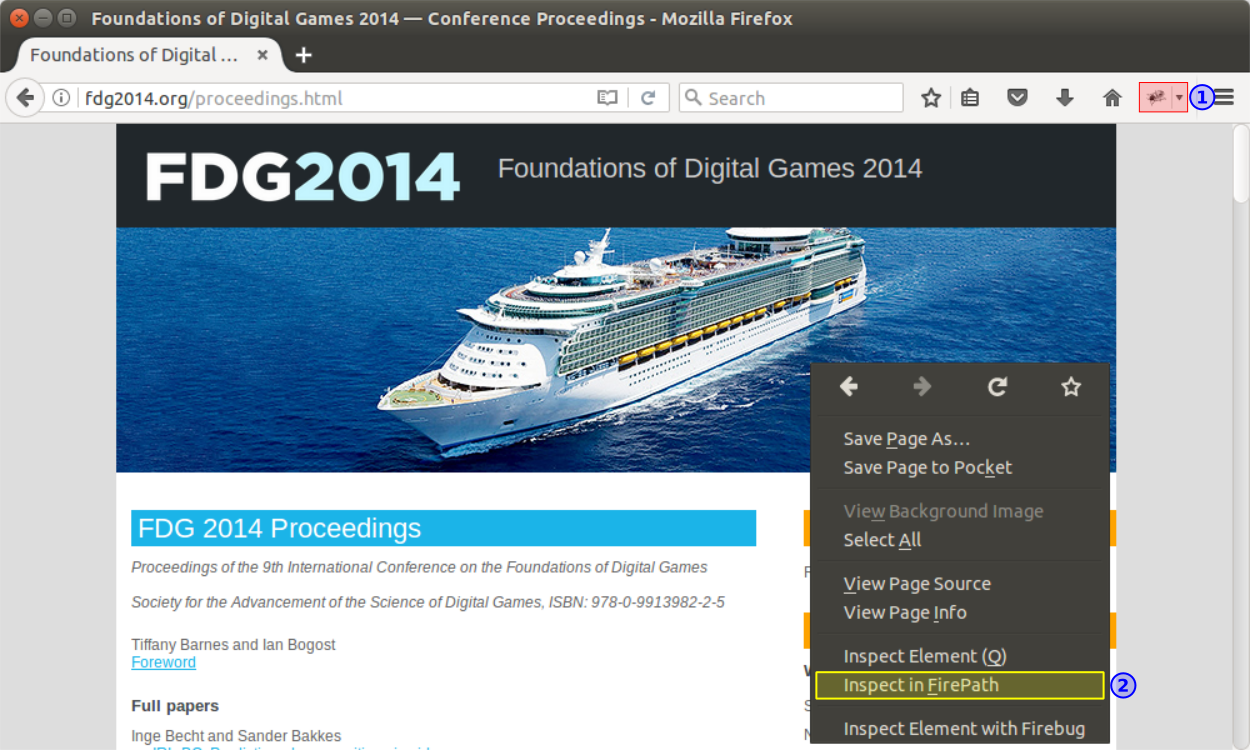}
\caption{\label{fdg2014-ff-context}Two easy ways of accessing the \emph{FirePath} tab in the \emph{Firebug} window: via (1) the context menu or (2) the Firebug button.}
\end{figure}

\begin{figure}[t]
	\centering
	\includegraphics[width=0.9\textwidth]{%
		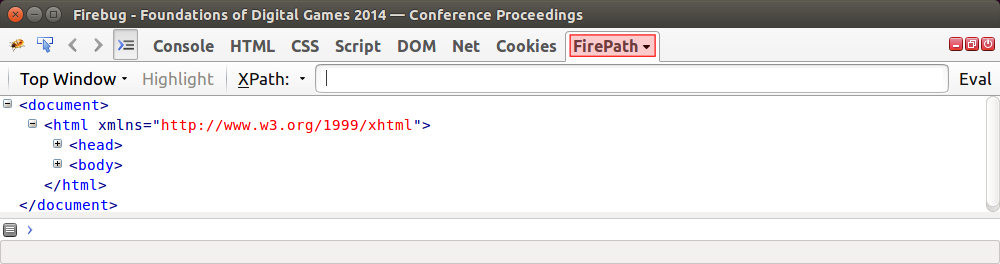}
	\caption{\label{fp-blank}An overview of the \emph{FirePath} tab in the \emph{Firebug} window.}
\end{figure}

The Firebug window which contains a tab labelled \emph{FirePath} (see Figure \ref{fp-blank}) can be opened by clicking on an icon with a grey bug in the tool bar (see Figure~\ref{fdg2014-ff-context}).
Opening FirePath can also be achieved by right-clicking an element of a web page and clicking the context-menu option \enquote{Inspect in FirePath} (see Figure~\ref{fdg2014-ff-context}).

\subsubsection{Usage of FirePath}


By selecting the FirePath tab and entering an XPath expression in the evaluation bar at the top of the Firebug window, both the validity and the selected nodes of an XPath query can be checked.
While the Firebug window unfolds the document tree and highlights the selected elements, a dashed border is added to the corresponding rendered elements in the browser window.
With the help of these add-ons, individual location paths to access the desired data can be written.

\begin{example}
In Figures~\ref{fdg2014-fp-mod} and \ref{fdg2014-fp-highlight},
\begin{figure}[t]
	\centering
	\includegraphics[width=11cm]{%
	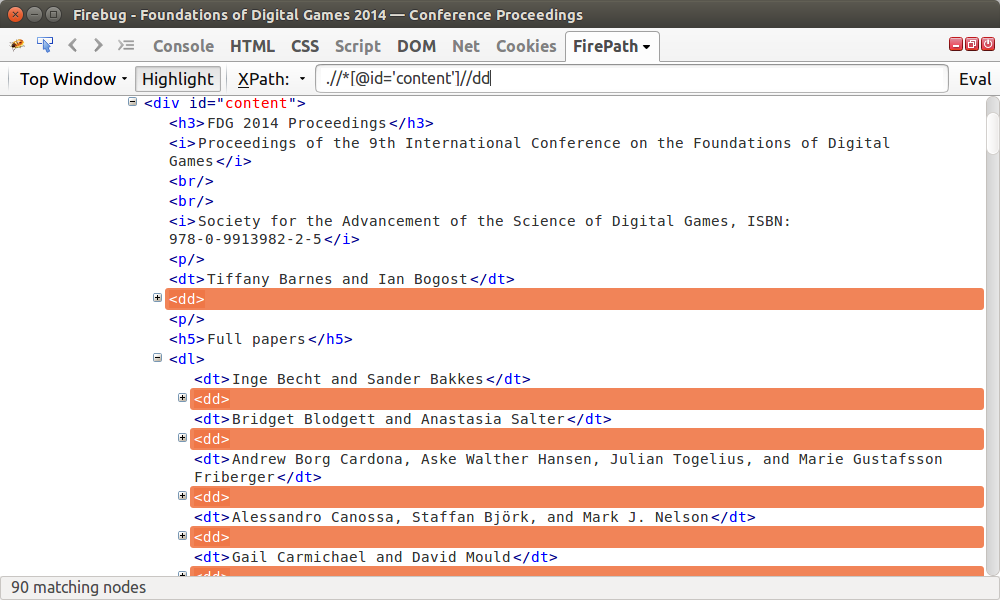}
	\caption{\label{fdg2014-fp-mod}A node set selected by XPath
	\lstinline[language=oxpath]{.//*[@id='content']//dd}
	to identify individual records on the FDG~2014 web page}
\end{figure}
\begin{figure}[t]
	\centering
	\includegraphics[width=11cm]{%
		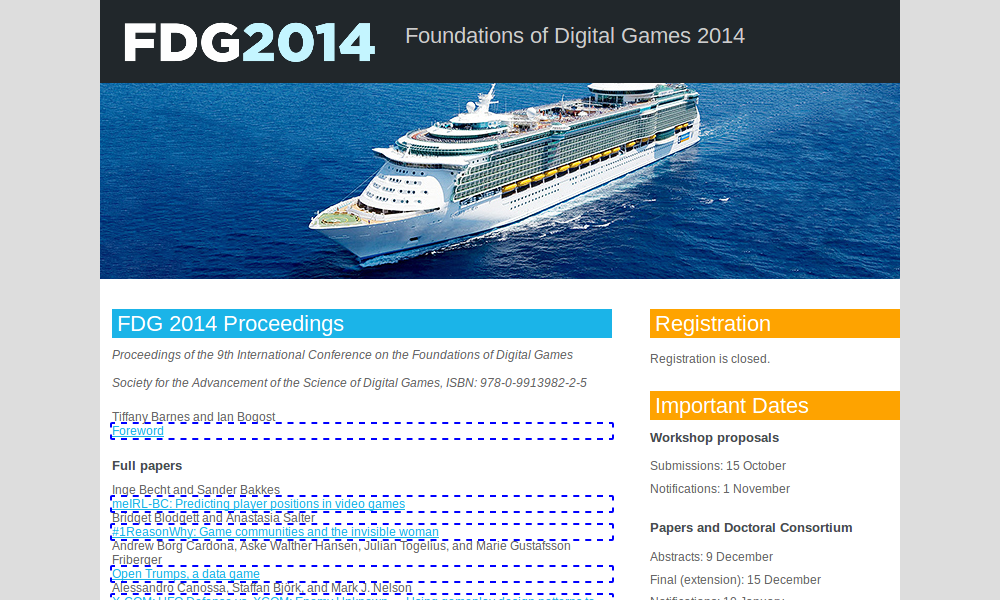}
	\caption{\label{fdg2014-fp-highlight}DOM nodes selected with the XPath expression
	\lstinline[language=oxpath]{.//*[@id='content']//dd}
	and highlighted by \emph{\emph{FirePath}} on the FDG~2014 web page.}
\end{figure}
we illustrate the result of evaluating the XPath expression
\lstinline[language=oxpath]{.//*[@id='content']//dd} with FirePath on a rendered web page FDG~2014.
\end{example}



%
%

\chapter{XPath}\label{xpath}



XPath is used to query XML documents descriptively. The ability to produce these descriptions requires an understanding of the basic concepts behind XPath expressions and their constituting parts. \emph{The W3C} introduces XPath with the following profile.

\blockquote{The primary purpose of XPath\index{XPath} is to address parts of an XML [..] document. In support of this primary purpose, it also provides basic facilities for manipulation of strings, numbers and booleans. [...] XPath operates on the abstract, logical structure of an XML document, rather than its surface syntax. XPath gets its name from its use of a path notation as in URLs for navigating through the hierarchical structure of an XML document.~\cite{w3c:xpath1.0-1999}}





\section{Concepts}\label{xpath:concepts}

\subsection{Node Trees}\label{xpath:meta:node_trees}

Similar to a tree of folders and files on a file system, \enquote{XPath models an XML document as a tree of nodes}. There are seven kinds of \emph{nodes}\index{XPath!Nodes} \cite{w3c:xpath1.0-1999}. Apart from \emph{namespace}, \emph{processing-instruction}, and \emph{comment nodes}, the remaining four \emph{node types} are encountered most commonly in XML documents (see Listing \ref{xpath:lst:node_trees} below):
The \emph{document node} is an artificial node with the root element as its only child ({\color{MaterialGreen}\texttt{results}}). All other elements of a document are considered \emph{element nodes} ({\color{MaterialGreen}\texttt{record}}, {\color{MaterialGreen}\texttt{date}}). The attributes of an element are regarded as \emph{attribute nodes} (\lstinline[language=xml]|type|, \lstinline[language=xml]|id|). Textual content in elements or attributes is represented by \emph{text nodes} ({\color{MaterialBlue}\texttt{2205-2222}}, {\color{MaterialBlue}\texttt{online}}).

\begin{caveat}
The names of element or attribute nodes such as {\color{MaterialGreen}\texttt{results}} or \lstinline[language=xml]|type| can be misleading in comparison with the examples of text nodes such as {\color{MaterialBlue}\texttt{2205-2222}}. Although element and attribute nodes are mentioned by their names only, selecting these nodes also includes their entire content.
\end{caveat}

All nodes exist in various relationships to each other. Each element and attribute node has one \emph{parent} ({\color{MaterialGreen}\texttt{record}} is the parent of {\color{MaterialGreen}\texttt{authors}}).
Element nodes may have zero, one or more \emph{children} ({\color{MaterialGreen}\texttt{title}}, {\color{MaterialGreen}\texttt{authors}}, {\color{MaterialGreen}\texttt{ee}}, {\color{MaterialGreen}\texttt{date}} and {\color{MaterialGreen}\texttt{pages}} are the children of each {\color{MaterialGreen}\texttt{record}}-node).
Nodes that have the same parent are called \emph{siblings} ({\color{MaterialGreen}\texttt{title}} and {\color{MaterialGreen}\texttt{authors}} are siblings). The parents of a node, their respective parents, etc. are called its \emph{ancestors} ({\color{MaterialGreen}\texttt{record}} and {\color{MaterialGreen}\texttt{results}} are the ancestors of {\color{MaterialGreen}\texttt{date}}). The children of a node, their respective children, etc. are called its \emph{descendants} (both {\color{MaterialGreen}\texttt{record}}-nodes and all their children are the descendants of {\color{MaterialGreen}\texttt{results}}). Nodes with no children or parents are also called \emph{atomic values}. Nodes and atomic values are also referred to as \emph{items}\cite{w3c:xpath2.0-2010}.

\begin{lstlisting}[%
  language=xml,
  caption=Example of an XML node tree,
  label=xpath:lst:node_trees
]
<results>
  <record id="1">
    <title>Adaptive bilateral control [...]</title>
    <authors>Zhang Chen, Bin Liang, Tao Zhang, [...]</authors>
    <ee>http://dx.doi.org/10.1017/S0263574714002847</ee>
    <date type="print">30 December 2014</date>
    <pages>2205-2222</pages>
  </record>
  <record id="2">
    <title>Design, kinematic and dynamic modeling [...]</title>
    <authors>Dan Zhang, Bin Wei</authors>
    <ee>http://dx.doi.org/10.1017/S0263574714002835</ee>
    <date type="online">23 December 2014</date>
    <pages>2186-2204</pages>
  </record>
</results>
\end{lstlisting}


\subsection{Path Expressions}\label{xpath:meta:path_expressions}

In the same way file paths identify files on a file system, following the \emph{location steps} of a \emph{path expression}\index{XPath!Path Expression} in XPath leads to a specific \emph{node} or \emph{node sets} in an XML document. Each location step is evaluated against the node set selected by the preceding step, where the current node is also referred to as the \textit{context node}. For the first step of a path, the document node usually serves as the basis for evaluation. As the following examples illustrate, the remaining steps navigate the node tree until the desired nodes are reached.

Based on the XML document in Listing \ref{xpath:lst:node_trees}, the following examples illustrate simple XPath expressions and the node sets they select. The selected nodes are presented below each of the presented XPath expressions.

\begin{minipage}{\linewidth}
\begin{lstlisting}[%
    language=oxpath,
    caption=Selection of simple nodes,
    label=xpath:lst:simple_nodes,
    linerange=1-1
  ]
/results/record/authors
\end{lstlisting}

\begin{lstlisting}[%
    language=xml,
  ]
(
  <authors>Zhang Chen, Bin Liang, Tao Zhang, [...]</authors>,
  <authors>Dan Zhang, Bin Wei</authors>
)
\end{lstlisting}
\end{minipage}

\begin{minipage}{\linewidth}
\begin{lstlisting}[%
    language=oxpath,
    caption=Selection of complex nodes,
    label=xpath:lst:complex_nodes,
    linerange=2-2
  ]
/results/record
\end{lstlisting}

\begin{lstlisting}[%
    language=xml,
  ]
(
  <record id="1">
    <title>Adaptive bilateral control [...]</title>
    <!--[...]-->
  </record>,
  <record id="2">
    <title>Design, kinematic and dynamic modeling [...]</title>
    <!--[...]-->
  </record>
)
\end{lstlisting}
\end{minipage}

\begin{minipage}{\linewidth}
\begin{lstlisting}[%
    language=oxpath,
    caption=Selection of attribute nodes,
    label=xpath:lst:attribute_nodes,
    linerange=3-3
  ]
/results/record/date/@type
\end{lstlisting}

\begin{lstlisting}[%
    language=xml,
  ]
(
  type="print",
  type="online"
)
\end{lstlisting}
\end{minipage}

\begin{minipage}{\linewidth}
\begin{lstlisting}[%
    language=oxpath,
    caption=Selection of text nodes,
    label=xpath:lst:text_nodes,
    linerange=4-4
  ]
/results/record/pages/text()
\end{lstlisting}

\begin{lstlisting}[%
    language=xml,
  ]
(
  2205-2222,
  2186-2204
)
\end{lstlisting}
\end{minipage}



%

\section{Syntax}\label{xpath:syntax}

Unlike the examples of the previous section, the individual steps of a path expression can be complex. A location step at least consists of an \emph{axis}, which defines the \emph{tree relationship} between the context node and the  nodes to be selected, a \emph{node test}, which specifies the desired nodes belonging to the selected axis further by name or type, and zero or more \emph{predicates} as another option to refine the selection (see Listing \ref{xpath:syntax}). A location path can be \emph{absolute} if it starts with a slash (\lstinline[language=oxpath]|/|), or \emph{relative} if it does not.

\begin{minipage}{\linewidth}
\begin{lstlisting}[%
  language=oxpath,
  caption=Simplified syntax of location steps,
  label=xpath:syntax
  ]
axisname::node-test[predicate]
\end{lstlisting}
\end{minipage}

\subsection{Axes}\label{xpath:syntax:axes}\index{XPath!Axes}%

Axes specify \enquote{the tree relationship between the nodes selected by the location step and the context node} \cite{w3c:xpath1.0-1999}. They largely rely on the existing node types and their relationships in the document tree.

\begin{longtable}{l X}
  \caption{List of XPath axes}\label{xpath:t:axes} \\
  \toprule\textbf{Axis name} & \textbf{Result} \\ \toprule\endfirsthead
  \toprule\textbf{Axis name} & \textbf{Result} \\ \toprule\endhead
  ancestor &  all ancestors (parent, grandparent, etc.) of the context node\\ \nopagebreak\midrule
  ancestor-or-self &  all ancestors (parent, grandparent, etc.) of the context node and the context node itself \\ \nopagebreak\midrule
  attribute &  all attributes of the context node \\ \nopagebreak\midrule
  child &  all children of the context node \\ \nopagebreak\midrule
  descendant &  all descendants (children, grandchildren, etc.) of the context node \\ \nopagebreak\midrule
  descendant-or-self &  all descendants (children, grandchildren, etc.) of the context node and the context node itself \\ \nopagebreak\midrule
  following &  everything after the closing tag of the context node \\ \nopagebreak\midrule
  following-sibling &  all siblings after the context node \\ \nopagebreak\midrule
  namespace &  all namespace nodes of the context node \\ \nopagebreak\midrule
  parent &  the parent of the context node \\ \nopagebreak\midrule
  \nopagebreak preceding &  all nodes that appear before the context node in the document, except ancestors, attribute nodes and namespace nodes, in \textbf{reverse order} \\ \nopagebreak\midrule
  preceding-sibling &  all siblings before the context node, in \textbf{reverse order} \\ \nopagebreak\midrule
  self & the context node \\ \bottomrule
\end{longtable}


\subsection{Node Tests}\label{xpath:syntax:node_tests}\index{XPath!Node Tests}%

Node tests may consist of a specific node name ({\color{MaterialGreen}\texttt{record}}, {\color{MaterialGreen}\texttt{date}}, etc.) or any of the pre-defined node tests \lstinline[language=oxpath]|comment()|, \lstinline[language=oxpath]|text()|, \lstinline[language=oxpath]|processing-instruction()|, or \lstinline[language=oxpath]|node()|.

\subsection{Predicates}\label{xpath:syntax:predicates}\index{XPath!Predicates}%

\enquote{A \emph{predicate} consists of an expression, called a \emph{predicate expression}, enclosed in square brackets. A predicate serves to filter a sequence, retaining some items and discarding others} \cite{w3c:xpath1.0-1999}. If the sub-expression inside a predicate returns an empty sequence for the current context node, that context node is excluded from the result sequence of the entire XPath expression. 
Predicates may be numeric. Numeric predicates specify the position of the selected node in the current node context. For instance, the expression in Listing \ref{xpath:lst:position} returns the first {\color{MaterialGreen}\texttt{record}}-child of {\color{MaterialGreen}\texttt{results}} (see Listing \ref{xpath:lst:node_trees}).

\begin{minipage}{\linewidth}
\begin{lstlisting}[%
    language=oxpath,
    caption=Selection by numeric predicate,
    label=xpath:lst:position,
    linerange=1-1,
  ]
/results/record[1]/date
\end{lstlisting}

\begin{lstlisting}[%
    language=xml,
  ]
(
  <date type="print">30 December 2014</date>
)
\end{lstlisting}
\end{minipage}

\begin{minipage}{\linewidth}
\begin{lstlisting}[%
    language=oxpath,
    caption=Selection by non-numeric predicate,
    label=xpath:lst:position,
    linerange=2-2,
  ]
/results/record/date[@type="online"]
\end{lstlisting}

\begin{lstlisting}[%
    language=xml,
  ]
(
  <date type="online">23 December 2014</date>
)
\end{lstlisting}
\end{minipage}


\subsection{Syntactic Sugar}\label{xpath:syntax:sugar}\index{XPath!SyntacticSugar}%
In addition to specify the desired axis explicitly in each step, it is also possible to write path expressions more concisely by using some syntactic sugar, a selection of which is presented below.

\begin{longtable}{l X}
  \caption{Various examples of syntactic sugar}\label{xpath:t:syntactic_sugar} \\
  \toprule\textbf{Path Expression} & \textbf{Description} \\ \toprule\endfirsthead
  \toprule\textbf{Path Expression} & \textbf{Description} \\ \toprule\endhead
  \emph{nodename} & Selects all nodes with the name \enquote{nodename} \\ \nopagebreak\midrule
  node() & Matches any node of any kind \\ \nopagebreak\midrule
  / &  step separator, implies child axis if no axis is specified \\ \nopagebreak\midrule
  // & Selects nodes in the same document as the context node that match the selection no matter where they are \\ \nopagebreak\midrule
  . & Selects the context node \\ \nopagebreak\midrule
  .. & Selects the parent of the context node \\ \nopagebreak\midrule
  @attribute-name & Selects attribute nodes with the name \texttt{attribute-name} \\ \nopagebreak\midrule
  * & Matches any element node \\ \nopagebreak\midrule
  @* & Matches any attribute node \\
  \bottomrule
\end{longtable}

\section{Evaluation}\label{xpath:evaluation}

The sets returned by the evaluation of an XPath expression are called \emph{sequences}. Sequences may be empty or consist of a single item only, but they cannot be nested. The contained items appear according to their order of occurrence in the document. Items can be duplicates and may belong to different node types.

\begin{minipage}{\linewidth}
\begin{lstlisting}[%
  language=oxpath,
  caption=Selection with an empty result sequence,
  label=xpath:lst:empty,
  linerange=1-1,
]
/results/record[3]
\end{lstlisting}

\begin{lstlisting}[%
  language=xml,
]
()
\end{lstlisting}
\end{minipage}

\section{Functions}\label{xpath:functions}

The following tables provide an overview of all functions originating from XPath 1.0 which are also supported by OXPath.
Most functions of the core function library in XPath 1.0 and some string functions from XPath 2.0 are supported and implemented by analogy with the corresponding W3C recommendations.
The core functions which differ from the implementation in those recommendations are listed along with the other OXPath-specific functions in Section \ref{sec:oxpath:spec-func} (s. pp. \pageref{sec:oxpath:spec-func}).
Each function is listed with its possible arguments.
The arguments in parentheses are required unless they have a trailing question mark.
For a more detailed explanation of individual functions, please, see the official documentation of XPath 1.0 by W3C (\cite{w3c:xpath1.0-1999}).
Functions such as {\color{MaterialOrange}\texttt{position()}} or {\color{MaterialOrange}\texttt{contains()}} are frequently used for node selection in OXPath expressions,
whereas {\color{MaterialOrange}\texttt{normalize-space()}} or  {\color{MaterialOrange}\texttt{replace()}} are commonly used for data extraction.

\begin{hint}
	If an argument is the single argument of a function and optional, it defaults to a node-set with the context node as its only member if omitted.
\end{hint}

\begin{longtable}{l l X}
  \caption{Node Set Functions from XPath 1.0
  }\label{xpath:t:functions:a} \\
  \toprule\textbf{Return Type} & \textbf{Function} & \textbf{Description} \\ \toprule\endfirsthead
  \toprule\textbf{Return Type} & \textbf{Function} & \textbf{Description} \\ \toprule\endhead





  number & last() & Returns the number equal to the context size from the expression evaluation context \\ \nopagebreak\midrule
  number & position() & Returns number equal to the context position from the expression evaluation context \\ \nopagebreak\midrule
  number & count(node-set) & Returns the number of nodes in the argument node-set \\ \nopagebreak\midrule

  string & local-name(node-set?) & Returns the local part of the expanded-name of the node in the argument node-set that is first in document order \\ \nopagebreak\midrule

  string & namespace-uri(node-set?) & Returns the namespace URI of the expanded-name of the node in the argument node-set that is first in document order (empty string except for element nodes and attribute nodes)  \\ \nopagebreak\midrule

  string & name(node-set?) & Returns a string containing a QName representing the expanded-name of the node in the argument node-set that is first in document order
      \\ \nopagebreak\bottomrule
\end{longtable}





\begin{longtable}{l l X}
  \caption{String Functions from XPath 1.0
  }\label{xpath:t:functions:b} \\
  \toprule\textbf{Return Type} & \textbf{Function} & \textbf{Description} \\ \toprule\endfirsthead
  \toprule\textbf{Return Type} & \textbf{Function} & \textbf{Description} \\ \toprule\endhead





  string & string(object?) & Converts an object to a string \\ \nopagebreak\midrule

  string & \makecell[tl]{concat\\~(string, string, string*)} & Returns the concatenation of its arguments \\ \nopagebreak\midrule
  boolean & starts-with(string, string) & Returns true if the first argument string starts with the second argument string, false otherwise \\ \nopagebreak\midrule
  boolean & contains(string, string) & Returns true if the first argument string contains the second argument string, false otherwise \\ \nopagebreak\midrule
  string & \makecell[tl]{substring\\~(string, number, number?)} & Returns the substring of the first argument starting at the position specified in the second argument with length specified in the third argument, continuing to the end of the string if omitted. \\ \nopagebreak\midrule

  number & string-length(string?) & Returns the number of characters in the string (see [3.6 Strings]). The argument defaults to the string-value of the context node if omitted. \\ \nopagebreak\midrule
  string & normalize-space(string?) & Returns the argument string with whitespace normalized by stripping leading and trailing whitespace and replacing sequences of whitespace characters by a single space. The argument defaults to the string-value of the context node if omitted. \\ \nopagebreak\midrule
  string & \makecell[tl]{translate\\~(string, string, string)} & Returns the first argument string with occurrences of characters in the second argument string replaced by the character at the corresponding position in the third argument string. For example, translate("bar","abc","ABC") returns the string BAr.
      \\ \nopagebreak\bottomrule
\end{longtable}

\begin{longtable}{l l X}
  \caption{Boolean Functions from XPath 1.0
  }\label{xpath:t:functions:c} \\
  \toprule\textbf{Return Type} & \textbf{Function} & \textbf{Description} \\ \toprule\endfirsthead
  \toprule\textbf{Return Type} & \textbf{Function} & \textbf{Description} \\ \toprule\endhead





  boolean & boolean(object) & Converts its argument to a boolean \\ \nopagebreak\midrule

  boolean & not(boolean) & Returns true if its argument is false, and false otherwise \\ \nopagebreak\midrule

  boolean & true() & Returns true \\ \nopagebreak\midrule
  boolean & false() & Returns false \\ \nopagebreak\midrule
  boolean & lang(string) & Returns true or false depending on whether the language of the context node as specified by xml:lang attributes is the same as or is a sublanguage of the language specified by the argument string.
    \\ \nopagebreak\bottomrule
\end{longtable}

\begin{longtable}{l l X}
  \caption{Number Functions from XPath 1.0
  }\label{xpath:t:functions:d} \\
  \toprule\textbf{Return Type} & \textbf{Function} & \textbf{Description} \\ \toprule\endfirsthead
  \toprule\textbf{Return Type} & \textbf{Function} & \textbf{Description} \\ \toprule\endhead





  number & number(object?) & Converts its argument to a number \\ \nopagebreak\midrule

  number & sum(node-set) & Returns the sum, for each node in the argument node-set, of the result of converting the string-values of the node to a number. \\ \nopagebreak\midrule
  number & floor(number) & Returns the largest (closest to positive infinity) number that is not greater than the argument and that is an integer. \\ \nopagebreak\midrule
  number & ceiling(number) & Returns the smallest (closest to negative infinity) number that is not less than the argument and that is an integer. \\ \nopagebreak\midrule
  number & round(number) & Returns the number that is closest to the argument and that is an integer. \\ \nopagebreak\bottomrule
\end{longtable}

\begin{longtable}{l l X}
  \caption{String Functions from XPath 2.0
  }\label{xpath:t:functions:e} \\
  \toprule\textbf{Return Type} & \textbf{Function} & \textbf{Description} \\ \toprule\endfirsthead
  \toprule\textbf{Return Type} & \textbf{Function} & \textbf{Description} \\ \toprule\endhead





  boolean & \makecell[tl]{matches\\~(string?, string, string)} & Returns true if the first argument matches the regular expression supplied as the second argument as influenced by the value of the third argument (if present), false otherwise.  \\ \nopagebreak\midrule
  string & \makecell[tl]{replace\\~(string?, string, string, string?)} & Returns the string that is obtained by replacing each non-overlapping substring of the first argument that matches the given pattern in the second argument with an occurrence of the string provided as the third argument as influenced by the value of the fourth argument (if present).
      \\ \nopagebreak\bottomrule
\end{longtable}

\section{References}\label{xpath:references}

\begin{itemize}
  \item \url{http://www.w3schools.com/xml/xpath_intro.asp}: A similar, brief introduction to XPath with some alternative examples
  \item \url{https://www.w3.org/TR/xpath/}: An extensive documentation of XPath 1.0 by W3C
\end{itemize}

\chapter{OXPath}

\OXPath is a web data extraction language, an extension of XPath 1.0 for interacting with web applications and extracting data from them.


\section{Design Goals and Implementation}

\OXPath pursues the following design goals~\cite{Furche2013-VLDB-OXPath}:

\textbf{Spirit of XPath.}
\OXPath follows the principles upon which XPath is built, in particular the use of a single, navigational expression, polynomial time evaluation (see~\cite{Furche2013-VLDB-OXPath} describing the complexity of the OXPath evaluation), and concise syntax.
Thus, OXPath expressions are path expressions just like plain XPath with concise syntax ensured by additional constructs, considered further in Section~\ref{sec:oxpath-constructs}.

\textbf{A tree-like result.}
The result of OXPath evaluation is a tree, that is generated with the use of \emph{extraction markers} (see Section~\ref{sec:extraction-marker}).
The structure of the result tree is implicitly specified by the path expression, thus, the shape of the extracted result reflects the shape of the expression.
This poses only a small limitation, as data on the Web are usually presented in a hierarchical way and OXPath implements all necessary XPath's axes (see Table~\ref{xpath:t:axes} on page~\pageref{xpath:t:axes}), making various navigation strategies through the DOM tree possible.

\textbf{Low memory.}
OXPath is suitable for large-scale data extraction, as its memory size remains small and constant throughout the execution regardless of the number of pages visited.
During the OXPath execution, the memory is occupied by the integrated web browser (i.e., Firefox) and the OXPath parser.
The memory footprint of the former mainly depends on the size of a web page (e.g., multimedia objects, fonts), while the latter usually occupies from 10 to 20 MB according to the analysis conducted in \cite[Section 6]{Furche2013-VLDB-OXPath}.

\textbf{Realistic interaction.}
OXPath allows the simulation of user actions to interact with the scripted multi-page interfaces of web applications.
The realistic simulation of user actions is possible due to the interaction with rendered states of web pages.
Thus, with its multi-page data model OXPath captures both page navigation and modifications to a page.
All simulated actions (see Section~\ref{sec:actions}) have a formal semantics and are specified declaratively with action types and context elements (i.e., current elements of the OXPath navigation path), such as the links to click on, or the form field to fill.

\textbf{Expressive and precise.}
OXPath inherits the precise selection capabilities of XPath (rather than heuristics for element selection) and, considerably increasing the language's expressiveness, extends them as follows:
\textbf{1.} OXPath enables navigation through page sequences (including multi-way navigation) and unbounded navigation sequences.
The former can be implemented with the use of the \emph{click} action (see Section~\ref{sec:actions}), for example, to follow multiple links from the same page.
The latter is realised by means of \emph{Kleene star} expression (see Section~\ref{sec:kleene-star}), for example, following \textit{next} links on a result page until there is no further such link.
\textbf{2.} OXPath enables the identification of data for extraction, which can be assembled into (hierarchical) records, regardless of its original structure (see Section~\ref{sec:extraction-marker} and \ref{sec:oxpathOutputTree}).

\textbf{Embeddable, standard API.}
OXPath is designed to integrate with other technologies, such as Java, XQuery, or Javascript. Following the spirit of XPath, it provides an API and hosts language to facilitate OXPath's interoperation with other systems (see Section~\ref{sec:integration}).

\section{OXPath-Specific Constructs}\label{sec:oxpath-constructs} 
The current version of OXPath (\OXPathWithVersion) extends XPath~1.0 with four main constructs:
1)~\emph{web page rendering} to fully render a web page in an integrated web browser,
2)~\emph{actions} to interact with the rendered interface of a web application,
3)~\emph{extraction markers} to select data to be extracted in a spirit of XPath expressions,
and
4)~\emph{Kleene star} to navigate through the set of pages or states of web applications with unknown extent.

\subsection{Web Page Rendering}\label{sec:oxpath:rendering}
Every OXPath expression always starts with a function call \lstinline[language=oxpath]{doc("URL")}, which specifies the URL of a web page to be loaded into the web browser.
OXPath waits until all relevant resources are loaded and rendered.
Sometimes a desired content is loaded after the web page has been rendered (i.e., after receiving the \textit{load} DOM event).
For example, if data loaded with the use of an AJAX request, this can require specification of additional waiting time in seconds as follows: \lstinline[language=oxpath]{doc("URL", [wait=T])}.
In this case, OXPath will wait {\color{MaterialRed}\texttt{T}} seconds before executing further steps of the OXPath expression.

\begin{remark}
	Before writing an OXPath expression, the {\color{MaterialRed}\texttt{wait}} parameter can be used to manually interact with a web page and ensure that all elements are correctly rendered in the integrated web browser and the OXPath, hence, can succeed in simulating user actions.
\end{remark}

\begin{example}
\begin{figure}
	\centering
	\includegraphics[width=12cm]{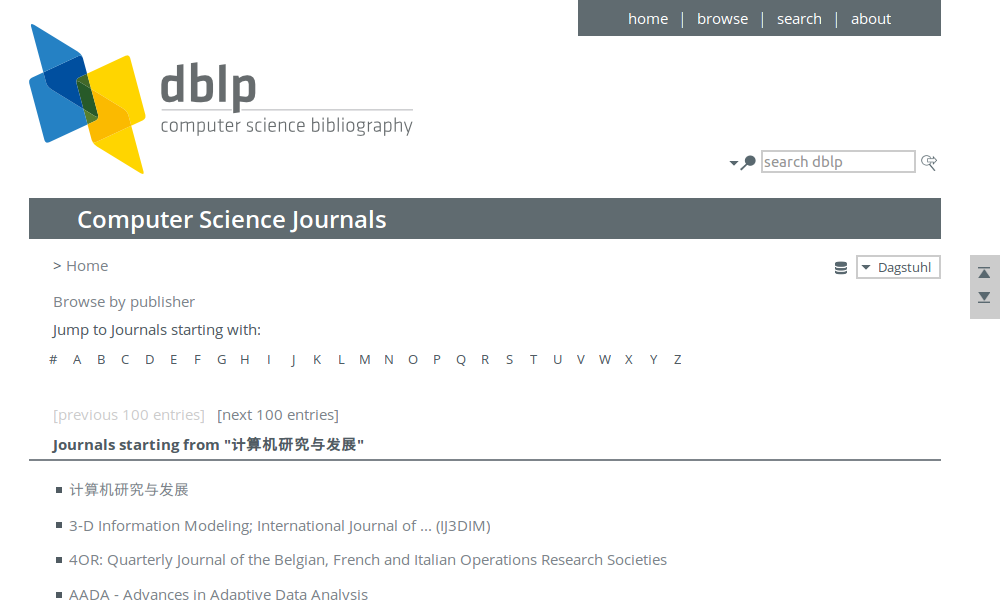}
	\caption{Computer science journals on dblp website\label{fig:oxpath:dblp-journal-pos1}}
\end{figure}
The following example loads dblp web page with the list of computer science journals (see Figure~\ref{fig:oxpath:dblp-journal-pos1}) and waits 60 seconds after the web page has been rendered.

		\begin{lstlisting}[language=OXPath]
doc("http://dblp.dagstuhl.de/db/journals/?pos=1", [wait=60])
		\end{lstlisting}
\end{example}

\subsection{Extraction Marker}\label{sec:extraction-marker}


Extraction markers are qualifiers, leveraged to build a tree-like structure with extracted data -- a result of the evaluation of an OXPath wrapper.
These markers extract data from DOM trees of traversed web pages and contribute to the output tree, which has a shape corresponding to the navigation path of the OXPath expression (see Section~\ref{sec:oxpathOutputTree}).
Each node in this tree can have a value of the context nodes (selected nodes in original DOM trees) modified with the use of XPath (see Tables~\ref{xpath:t:functions:a}--\ref{xpath:t:functions:e} on pages~\pageref{xpath:t:functions:a}--\pageref{xpath:t:functions:e}) or OXPath-specific (see Tables~\ref{oxpath:t:functions:a}--\ref{oxpath:t:functions:c} on pages ~\pageref{oxpath:t:functions:a}--\pageref{oxpath:t:functions:c}) functions.
An extraction marker is represented as
{\color{MaterialGreen}\texttt{:<name>}}
or
{\color{MaterialGreen}\texttt{:<name=value-exp>}},
in which \texttt{value-exp} is a constant, an XPath expression, or an XPath or OXPath-specific function.
The evaluation of \texttt{value-exp} should return string, numeric or boolean value.
It is important to mention that extraction markers may not occur inside of other extraction markers, function, or operator arguments.

The structure of the tree is not bound to the structure of the page due to the variety of axes which can define the navigation strategy through the DOM tree (see Table~\ref{xpath:t:axes} on page~\pageref{xpath:t:axes}).
Predicates are used to specify parent-child relations.
For example, \lstinline[language=oxpath]{//a:<o1>[.//b:<o2>]}
sets the parent-child relation between output nodes \texttt{o1} and \texttt{o2}.

The name of the extraction marker (that results into the \emph{output node}) is a sequence of characters, starting with letter or underscore and continuing with letters, digits, underscore, and hyphens.
Thus, the allowed set of name of extracted nodes is a subset of names allowed by the W3C recommendation~\cite[Section 2.3]{w3c-xml1.0-2008}.

\begin{example} \label{exmpl:extractionMarkers}
	The following OXPath wrapper extracts the title of the dblp web page (see Figure~\ref{fig:oxpath:dblp-journal-pos1}), ``Computer Science Journals'', enclosed by the tag
	{\color{MaterialGreen}\texttt{title}},
	and names of journals in the content listed.
	The \texttt{title} and \texttt{name} nodes are serialised as siblings.

	\begin{lstlisting}[language=oxpath]
doc("http://dblp.dagstuhl.de/db/journals/?pos=1")
  //header[@class~="headline"]:<title=string(./h1)>
  /..//div[@class="hide-body"]//a:<name=string(.)>
	\end{lstlisting}

	The OXPath-specific operator \textasciitilde{}\texttt{=} ensures that the right operand is contained in the left operand as a word.
	The XML serialization of the result looks as follows.
	\begin{lstlisting}[language=xml]
<results>
  <title>Computer Science Journals</title>
  ...
  <name>3-D Information Modeling; International Journal of  ... (IJ3DIM)</name>
  <name>4OR: Quarterly Journal of the Belgian, French and Italian Operations Research Societies</name>
  ...
</results>
	\end{lstlisting}

	The parent-child relation (and, therefore, nesting of tags in the XML serialization) can be achieved with the use of predicates.
	For example, the expression below sets the parent-child relation between extraction nodes
	{\color{MaterialGreen}\texttt{journals}}
	and
	{\color{MaterialGreen}\texttt{journal}}
	as well as between
	{\color{MaterialGreen}\texttt{journal}}
	and the pair
	{\color{MaterialGreen}\texttt{name}}--{\color{MaterialGreen}\texttt{url}}.

\begin{lstlisting}[language=OXPath]
doc("http://dblp.dagstuhl.de/db/journals/?pos=1")
  //header[@class~="headline"]:<title=string(./h1)>
  :<journals>
    [./..//div[@class="hide-body"]//a:<journal>
      [.:<name=string(.)>:<url=string(@href)>]
    ]
\end{lstlisting}

	This wrapper produces the following XML.
\begin{lstlisting}[language=xml]
<results>
  <title>Computer Science Journals</title>
  <journals>
    <journal>
      <name>3-D Information Modeling; International Journal of  ... (IJ3DIM)</name>
      <url>http://dblp.dagstuhl.de/db/journals/ij3dim</url>
    </journal>
    <journal>
      <name>4OR: Quarterly Journal of the Belgian, French and Italian Operations Research Societies</name>
      <url>http://dblp.dagstuhl.de/db/journals/4or</url>
    </journal>
    ...
  <journals>
</results>
\end{lstlisting}

\end{example}

%

%

\subsection{Actions}\label{sec:actions}


For simulating user interactions such as clicks or typing, OXPath introduces \emph{actions} (see Table~\ref{tbl:oxpathActions}), which constitute \emph{contextual} and \emph{absolute action steps}.
Contextual action steps have a form
{\color{MaterialRed}\texttt{\{action\}}}.
They keep the current context node after the evaluation, i.e., the subsequent OXPath step is always evaluated in regards to a node that is current for the contextual action.
Absolute action steps, as in
{\color{MaterialRed}\texttt{\{action /\}}},
set the root of the DOM tree as a new context node.

Contextual actions can be used in case if the web page does not change or, to be precise, if the path from the root of the DOM tree till the current node stays the same%
\footnote{This fact is mainly related to the evaluation of the \emph{action-free prefix}~\cite{Furche2013-VLDB-OXPath} of the performed action, which is obtained from the OXPath expression starting at the previous absolute action by dropping all intermediate contextual actions and extraction markers.}.
Absolute actions are useful in the following cases:
(i) the web page changes considerably or (ii) subsequent OXPath steps are not related to the current context node.
Regarding (i), the interaction might involve navigation to a new web page or invocation of an AJAX request, altering the current context (e.g., by removing it).
Concerning (ii), it might be the case, that after applying the action, the simplest XPath expression to select the next relevant DOM node is from the root.

It is worth mentioning that actions may not occur in other extraction markers, function, or operator arguments.
\begin{longtable}{l l X}
  \caption{OXPath actions\label{tbl:oxpathActions}} \\
  \toprule\textbf{Name} & \textbf{Example} & \textbf{Description} \\ \toprule\endfirsthead
  \toprule\textbf{Name} & \textbf{Example} & \textbf{Description} \\ \toprule\endhead
  Click & \texttt{click} &  Click on a current element.\\ \nopagebreak\midrule
  Next click   & \texttt{nextclick} & Click on an element and return a context node only if the resulting web page has the URL which has not been observed earlier with the next click action.
  This action is useful in conjunction with the Kleene star (see Section~\ref{sec:kleene-star}).\\ \nopagebreak\midrule
  Click with change   & \texttt{clkwithchange} &  Click on an element and return a context node only if the resulting web page differs from the initial.
  This action is useful in conjunction with the Kleene star.\\ \nopagebreak\midrule
  Type   & \texttt{"text"} & Simulate typing user action on the current element.\\ \nopagebreak\midrule
  Press enter  & \texttt{pressenter} &  Press enter key on the current element.\\ \nopagebreak\midrule
  Submit & \texttt{submit} & Submit the current web form.\\ \nopagebreak\midrule
  Focus   & \texttt{focus} & Throw the focus event on the current element.\\ \nopagebreak\midrule
  Mouse over   & \texttt{mouseover} & Throw the mouse over event on the current element.\\ \nopagebreak\midrule
%
%
  Move to a frame   & \texttt{movetoframe} & If the current node is either \texttt{iframe} or \texttt{frame} and the \texttt{src} attribute is specified, navigate to a web page with the URL specified by \texttt{src}.
  It throws \texttt{WebAPIRuntimeException} if the element is not an \texttt{iframe} or \texttt{frame}, and if \texttt{src} is not defined.\\ \nopagebreak\midrule
  Enter a frame   & \texttt{enterframe} & If a current node is either \texttt{iframe} or \texttt{frame} and the \texttt{src} attribute is specified, change the context node to the \texttt{html} node of the DOM tree of the frame node.\\ \bottomrule
\end{longtable}

\begin{caveat}
	Use absolute actions steps if their evaluation either loads a new web page or considerably changes the content of a web page.
\end{caveat}

\begin{remark}
	Actions such as clicking and typing simulate real user interaction, which involves scrolling to the relevant element until it is presented in the browser's viewport, if it is not there, and simulating the \textit{mouse over} and \textit{focus} events.
\end{remark}

\begin{figure}[h]
	\centering
	\includegraphics[width=12cm]{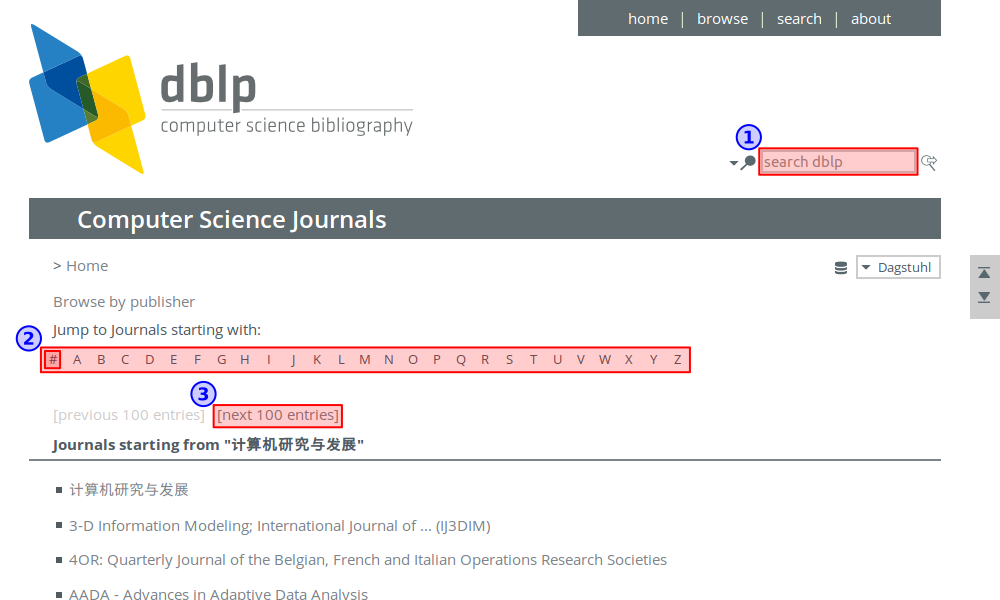}
	\caption{Computer science journals on DBLP website\label{fig:oxpath:dblp-journal-pos1:firstLetterNavigation}}
\end{figure}

\begin{example}
	For keyword-based searching (option 1 in Figure~\ref{fig:oxpath:dblp-journal-pos1:firstLetterNavigation}), we can type into the search field with the following expression:
	\begin{lstlisting}[language=OXPath]
doc("http://dblp.dagstuhl.de/db/journals/?pos=1")
  //div[@id="search"]/form[1]/field()[1]/{"Very Large Data Bases (VLDB)"}
	\end{lstlisting}
This expression inputs ``Very Large Data Bases (VLDB)'' into the search field.
\end{example}

In most cases, clicking is the main mean for navigating pages.
OXPath supports \emph{multi-way navigation} allowing the traversal through different navigation paths which can be represented as a tree, that is traversed in a depth-first fashion.
Forks, representing navigational alternatives, are mainly the result of multiple node selections.
Thus, all alternative paths are traversed sequentially and the browser history is used for the back locomotion through the sequence of visited pages.

\begin{example}\label{exmpl:click-action}
	To navigate through all lists of  journals from the dblp website we can navigate the alphabetical list (option 2 in Figure~\ref{fig:oxpath:dblp-journal-pos1:firstLetterNavigation}) with the following OXPath expression.
	\begin{lstlisting}[language=OXPath]
doc("http://dblp.dagstuhl.de/db/journals/?pos=1")
  //nav[@class="prefix-index"]//div[@class="body"]/a/{click/}
	\end{lstlisting}
In this example,
\lstinline[language=oxpath]{//nav[@class="prefix-index"]//div[@class="body"]/a}
selects links to different catalogues of journals, and the subsequent click action forces the browser to navigate to the corresponding list of journals.
A set of catalogue links along with click interaction in this case, mould out a multi-way navigation, that is illustrated in Figure~\ref{fig:exmpl-multiway-nav}.
\begin{figure}
	\centering
	\includegraphics[width=9cm]{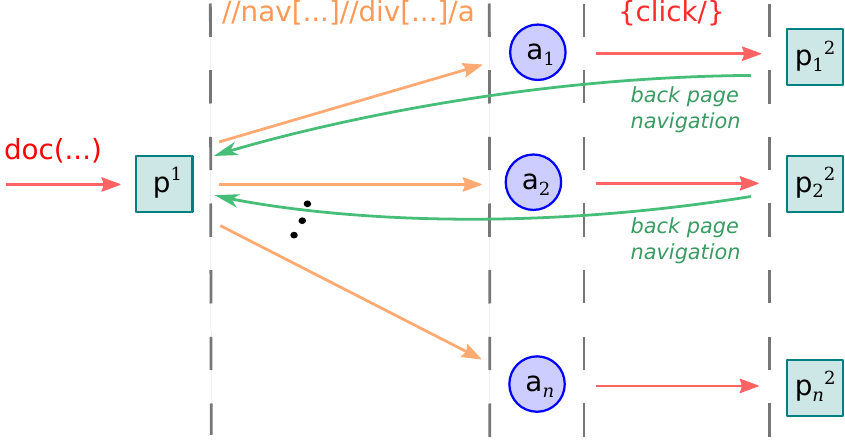}
	\caption{\label{fig:exmpl-multiway-nav}Graphical example of the multi-way navigation for extracting names of journals from the dblp website (see Example~\ref{exmpl:click-action})}
\end{figure}
OXPath sequentially interacts with all catalogue links $a_i$ to navigate from the original page $\text{p}^1$ to corresponding pages $\text{p}^2_i$.
It starts with $a_1$ leading to $\text{p}^2_1$.
A back navigation through the browser history is utilised to get back to the original $\text{p}^1$ page to navigate from $\text{p}^2_i$ to $\text{p}^2_{i+1}$.
\end{example}

Interactive and AJAX-rich interfaces require a certain amount of time for user actions to be processed by the JavaScript engine of the web browser.
In spite of the fact that the OXPath engine analyses DOM events and changes on a web page to identify the right moment for applying the next OXPath step, sometimes it is necessary to manually set specific time interval.
The delay between the current action step and the next OXPath step can be specified in seconds as follows:
{\color{MaterialRed}\texttt{\{action [wait=T]\}}}
and
{\color{MaterialRed}\texttt{\{action [wait=T]/\}}}
for contextual and absolute action steps, respectively, in which
{\color{MaterialRed}\texttt{T}}
is the time interval.

\begin{hint}
	To force the OXPath engine to wait a certain amount of seconds after the evaluation of an action step or the load of a web page, use wait statements:
	{\color{MaterialRed}\texttt{\{action [wait=T]\}}}
	and
	{\color{MaterialRed}\texttt{\{action [wait=T]/\}}}
	for action steps and
	\lstinline[language=oxpath]{doc("URL", [wait=T])}
	for rendering a web page.
\end{hint}

\subsection{Kleene Star}\label{sec:kleene-star}

Kleene star allows applying the same OXPath sub-expression multiple times.
There are two types of Kleene star:
\emph{unbounded}
{\color{MaterialPurple}\texttt{(}}
{\color{MaterialOrange}\texttt{\textit{OXPath expression}}}
{\color{MaterialPurple}\texttt{)*}}
for the unknown extent (i.e., the multiplicity is limited by the input sources) and \emph{bounded}
{\color{MaterialPurple}\texttt{(}}
{\color{MaterialOrange}\texttt{\textit{OXPath expression}}}
{\color{MaterialPurple}\texttt{)*}}
{\color{MaterialRed}\texttt{\{m,n\}}}
with specified lower \texttt{m} and upper \texttt{n} bounds (where \texttt{m} and \texttt{n} are non-negative integers and \texttt{m}$\leq$\texttt{n}).
The unbounded Kleene star is evaluated in a spirit of ``zero or more'' times, i.e., it is omitted in the 0-iteration, however, all subsequent OXPath steps are evaluated on each iteration.
The \emph{bounded Klenee star} is evaluated \texttt{n}$-1$ times at most (the 0-iteration is always skipped), however, all subsequent steps are processed starting from the iteration \texttt{m}, i.e., they are evaluated \texttt{n}$-$\texttt{m}$+1$ times.
The lower bound set to $1$ (i.e., \texttt{m}$=1$) means that a Kleene Star expression and subsequent OXPath steps will be omitted on the 0-iteration and will be evaluated starting with iteration 1.
In contrast, if \texttt{m}$=0$, then Kleene star will not take part in the first iteration, and the subsequent steps, therefore, will be processed one time more.
If evaluation of Kleene start results in empty set, its evaluation and evaluation of subsequent OXPath steps are terminated.

There are some restrictions worth mentioning.
The XPath expression in Kleene star is evaluated against the root of the current DOM tree regardless of the current context node and all subsequent OXPath steps are evaluated in regard to the context obtained recently.
Furthermore, only one Kleene star expression is allowed in an OXPath wrapper.

The evaluation of Kleene star expression can also be explained as a loop.
Algorithm~\ref{alg:kleeneestarloop} demonstrates the execution both of the unbound (the \textit{unboundKleneeStar} function) and bound (the \textit{boundKleneeStar} function) versions of Kleene star.
\begin{algorithm}[t]
	\caption{An algorithmic interpretation of Klenee star}
	\label{alg:kleeneestarloop}
	\DontPrintSemicolon
	\Fn{\textnormal{\textit{unboundedKleneeStar}}($kleeneStarExpr$, $subseqSteps$)} {
		$i \leftarrow 0$; $success \leftarrow True$

		\While{$success = True$}{
			\lIf{$i>0$}{
				$success \leftarrow evaluate(kleeneStarExpr)$
			}
			\lIf{$success = True$}{
				$evaluate(subseqSteps)$
			}
			$i \leftarrow i+1$
		}
	}

	\Fn{\textnormal{\textit{boundedKleneeStar}}($kleeneStarExpr$, $subseqSteps$, $m$, $n$)} {
		$i \leftarrow 0$; $success \leftarrow True$

		\While{$success = True$ and $i \leq n$}{
			\lIf{$i>0$}{
				$success \leftarrow evaluate(kleeneStarExpr)$
			}
			\lIf{$success = True$ and $i \geq m$}{
				$evaluate(subseqSteps)$
			}
			$i \leftarrow i+1$
		}
	}
\end{algorithm}

One of the most frequent use cases is the navigation through search result pages by clicking the ``next page'' link (see Example~\ref{exmpl:kleenestart-nextpage}).

\begin{hint}
	Kleene star can be a very useful mean if the iterative extraction over a set of similar pages is necessary and simulation of user interaction for the transition between pages is required.
\end{hint}

\begin{example}\label{exmpl:kleenestart-nextpage}
	Continuing Example~\ref{exmpl:click-action}, the list of result pages can also be navigated by sequentially clicking the link, ``[next 100 entries]'' (option 3 in Figure~\ref{fig:oxpath:dblp-journal-pos1:firstLetterNavigation} on page~\pageref{fig:oxpath:dblp-journal-pos1:firstLetterNavigation}).
	This can be realised with actions \texttt{nextclick} or \texttt{clkwithchange} combined with Kleene star.
	The next click action (\texttt{nextclick}) will be successfully evaluated until we navigate to a web page with the URL visited earlier.
	In our case, until the next link becomes inactive and we stay on the same webpage.
	In contrast, click-with-change (\texttt{clkwithchange}) stops the evaluation, if the web page content is the same as it was before.

	This scenario is implemented in the following OXPath expression with the use of the \texttt{nextclick} absolute action and the extraction marker \texttt{name} for journals.
		\begin{lstlisting}[language=OXPath]
doc("http://dblp.dagstuhl.de/db/journals/?pos=1")/
  ?!(!?//div[@id="browse-journals-output"]/p[1]/a[string(.)="[next 100 entries]"]/{nextclick/}?!)*!?
  //div[@class="hide-body"]//a:<name=string(.)>
		\end{lstlisting}

	This example can also be implemented with the \texttt{click} action, however, the termination state should be specified.
	It can either be specified by the boundaries of Kleene star, equal to the number of required iterations, or, which is more practical, by the XPath expression within the Kleene star expression that selects an empty set when there are no more result pages, e.g.:
	
\begin{lstlisting}[language=OXPath]
?!(!?//div[@id="browse-journals-output"]/p[1]/a[string(.) = "[next 100 entries]" and string (@class) != "disabled"]/\{click/\}?!)*!?
\end{lstlisting}

	
	In this example, the statement
	\lstinline[language=oxpath]{string(@class)!="disabled"}
	makes the Kleene star expression ``invalid'' on the last page, when the link ``[next 100 entries]'' is deactivated with the class ``disabled''.
	In this case, when Kleene star returns empty set, its evaluation and evaluation of subsequent OXPath steps are terminated.
\end{example}

\begin{caveat}
	When using Kleene star, do not forget to take care of termination states.
	They can be explicitly defined by the bounds of Kleene star, implicitly by the peculiarities of used action (e.g., \texttt{nextclick} with its evaluation conditions), or by the plain XPath expression within the Kleene star which selects the empty set at a proper moment.
\end{caveat}

\begin{example}
	The extraction marker can be incorporated into the Kleene star expression.
	The following OXPath wrapper inserts
	{\color{MaterialGreen}\texttt{kleenestareval}} tag into the result tree each time Kleene star is evaluated.
		\begin{lstlisting}[language=OXPath]
doc("http://dblp.dagstuhl.de/db/journals/?pos=1")/
?!(!?//div[@id="browse-journals-output"]/p[1]/a[string(.)="[next 100 entries]"]/{nextclick/}/.:<kleenestareval>?!)*!?
//div[@class="hide-body"]//a:<name=string(.)>
		\end{lstlisting}
\end{example}

\section{Some Restrictions and Peculiarities}

\begin{enumerate}
	\item Actions and extraction markers may not occur inside of other extraction markers, function, or operator arguments.
	\item The value of an extraction marker must yield a scalar value: string, number or boolean.
	\item Only one Kleene-starred expression is allowed.
	\item No page identity. OXPath does not manage page ``identity'': If two links lead to the same URL, OXPath considers the pages reached by clicking on those links as distinct.
	This avoids issues with server state where the same URL returns different results at different times or points in an interaction.
	It also avoids the need to maintain pages in OXPath in case they are later encountered again.
	\item No back. OXPath does not allow the reverse (or ``back'') navigation over DOM trees of pages.
	That is, plain XPath expressions are always evaluated for the current DOM tree.
	Once we have moved from page A to B, there is no way to explicitly define back-navigation to A.
	This is a limitation as it (together with the lack of variables) prohibits a class of wrappers that refer back to values encountered on earlier pages.
	However, it is essential to maintain the low memory profile of OXPath.
\end{enumerate}

\section{OXPath-Specific Functions}\label{sec:oxpath:spec-func}

OXPath implements standard functions of XPath~1.0 and XPath~2.0 listed in Tables~\ref{xpath:t:functions:a}--\ref{xpath:t:functions:e} on pages~\pageref{xpath:t:functions:a}--\pageref{xpath:t:functions:e} and additional functions introduced in Tables~\ref{oxpath:t:functions:a}--\ref{oxpath:t:functions:c}.
Note that although some of these functions, e.g. \texttt{substring-before} or \texttt{substring-after} are already part of the XPath 1.0 specification, they are listed here because they have been reimplemented within OXPath with the possibility for their further extension.

\begin{hint}
	If an argument is the single argument of a function and it is optional, then in case it is omitted, the function is evaluated over the current context node.
\end{hint}

\begin{longtable}{l l X}
  \caption{Functions to Access Node Content in OXPath
  }\label{oxpath:t:functions:a} \\
  \toprule\textbf{Return Type} & \textbf{Function} & \textbf{Description} \\ \toprule\endfirsthead
  \toprule\textbf{Return Type} & \textbf{Function} & \textbf{Description} \\ \toprule\endhead

  string & to-xml(node?) & Returns a string serialization of the XML representation of the specified node. \\ \nopagebreak\midrule
  string & innerhtml(node?) & Returns the \texttt{innerHTML} property of the specified node.\\ \nopagebreak\midrule
  string & outerhtml(node?) & Returns the \texttt{outerHTML} property of the specified node. \\ \nopagebreak\midrule
  string & page-content() & Returns a string serialization of HTML of the current state of a web page.
      \\ \nopagebreak\bottomrule
\end{longtable}


\begin{longtable}{l l X}
  \caption{String Functions in OXPath
  }\label{oxpath:t:functions:b} \\
  \toprule\textbf{Return Type} & \textbf{Function} & \textbf{Description} \\ \toprule\endfirsthead
  \toprule\textbf{Return Type} & \textbf{Function} & \textbf{Description} \\ \toprule\endhead

  string & \makecell[tl]{string-join\\~(node-set, string, string?)} & Returns a string created by concatenating the string values of the members of the first argument with the second argument as a separator.
  The third argument is optional and specifies the procedure for obtaining the textual representation of nodes,
  accepting the following values: ``to-xml'', ``normalize-space'', ``innerhtml'', and ``outerhtml''.
      \\ \nopagebreak\midrule
  string & \makecell[tl]{substring-before\\~(string, string, number?)} & Returns the substring of the first argument string that precedes the first occurrence of the second argument string in the first argument string, or the empty string if the first argument is not retrieved.
  The third parameter is optional and specifies the ordinal number of the occurrence of the second argument to be considered as a separator.
  (It is 1 by default.)
  For example, \texttt{substring-before(., "separator", 5)} returns the substring of the first argument that starts from the beginning of the string and ends just before the 5th occurrence of the substring ``separator''.
      \\ \nopagebreak\midrule
  string & \makecell[tl]{substring-after\\~(string, string, number?)} & Returns the substring of the first argument string that follows the first occurrence of the second argument string in the first argument string, or the empty string if the first argument is not retrieved.
  The third parameter is optional and specifies the ordinal number of the occurrence of the second argument to be considered as a separator.
  (It is 1 by default.)
  For example, \texttt{substring-after(., "separator", 5)} returns the substring of the first argument that starts just after the 5th occurrence of the substring ``separator'' and ends at the end of the string.
      \\ \nopagebreak\midrule
  string & \makecell[tl]{substring-befor-reverse\\~(string, string, number?)} & Returns a substring as in \texttt{substring-before} but, in contrast, the search of the ``separator'' is conducted in a reverse direction, i.e., from the last to the first character of the string. \\ \nopagebreak\midrule
  string & \makecell[tl]{substring-after-reverse\\~(string, string, number?)} & Returns a substring as in \texttt{substring-after} but, in contrast, the search of the ``separator'' is conducted in a reverse direction, i.e., from the last to the first character of the string. \\ \nopagebreak\midrule
  string & \makecell[tl]{substring-reverse\\~(string, string, number?)} & Returns a substring as XPath's \texttt{substring} function but, in contrast, the input string is reverted.
      \\ \nopagebreak\bottomrule
\end{longtable}


\begin{longtable}{l l X}
  \caption{Other Additional Functions in OXPath
  }\label{oxpath:t:functions:c} \\
  \toprule\textbf{Return Type} & \textbf{Function} & \textbf{Description} \\ \toprule\endfirsthead
  \toprule\textbf{Return Type} & \textbf{Function} & \textbf{Description} \\ \toprule\endhead




  number & \makecell[tl]{opex:minus\\~(number, number)} & Subtracts the second argument from the first one. \\ \nopagebreak\midrule
  number & \makecell[tl]{opex:plus\\~(number, number)} & Adds the first and the second argument. \\ \nopagebreak\midrule
  number & \makecell[tl]{jaro-wrinkler\\~(object, object)} & Returns the similarity between the two arguments (a DOM node or a string) in terms of Jaro-Winkler distance.
  The value is in the range from 0 to 1, the higher the similarity the higher the value.
      \\ \nopagebreak\midrule

  string & current-url() & Returns the URL of the current web page.\\ \nopagebreak\midrule
  string & qualify-url(string) & Returns the absolute URL for the relative URL specified in the argument based on the URL of the current web page. \\ \nopagebreak\midrule

  boolean & is-visible(object?) & Returns \texttt{true} if the current element has been visualized by the web browser and has a non-empty area. \\ \nopagebreak\midrule
  string & dom-property(string) & Returns the property specified by the argument for the current DOM node. For example, \texttt{dom-property("nodeType")} returns the type of a DOM node. \\ \nopagebreak\midrule

  string & now() & Creates a timestamp in a form \texttt{yyyy-MM-dd'T'HH:mm:ss'Z'}.\\ \nopagebreak\midrule
  string & screenshot() & Loads a web page with the URL of a current web page into a new window and makes a screenshot. It returns the path to the file created. The path has the following form \texttt{\textit{host\_timestamp}}, in which all characters of \texttt{\textit{host\_timestamp}} matching the regular expression ``\texttt{\textbackslash{}W}'' are replaced by ``\texttt{\_}'', \texttt{\textit{timestamp}} has the format ``\texttt{ddMMM\_HHmmss}''.
      \\ \nopagebreak\bottomrule
\end{longtable}


\section{Syntactic Sugar and Operators}\label{sec:oxpath:syntsugar-op}

OXPath implements additional binary operators:
\textbf{1)} \textasciitilde{}\texttt{=} returns \texttt{true} if the right operand is within a list of space-separated values of the left operand;
\textbf{2)} \texttt{\#=} returns \texttt{true} if the right operand is a substring of the left operand (it is a syntactic sugar for the \texttt{substring} function).

The function \texttt{is-invisible()} is a syntactic sugar for the expression \texttt{not(is-visible())}.

One of the most useful constructs extending predicates is an optional predicate.
It is of the form
\texttt{{\color{MaterialOrange}[? \textit{\text{OXPath expression}} ]}}.
This expression equals
\texttt{{\color{MaterialOrange}[ \textit{\text{OXPath expression} }  or true()]}}
and is effective for optional attributes in records to be extracted.


\begin{hint}
	Optional predicates
	\texttt{{\color{MaterialOrange}[? \textit{\text{OXPath expression}} ]}}
	are useful for extracting data records with optional attributes.
\end{hint}

\section{OXPath Output Tree}\label{sec:oxpathOutputTree}

An OXPath output tree is generated as the result of OXPath evaluation.
It is constructed ``on-the-fly'' from the extraction markers and is streamed out to the underlying output handlers, converting the tree into a desired format.
The tree contains three main types of nodes: the root, a record node, and attribute node.
A record node can contain other record or attribute nodes as its children, while attribute nodes, conveying extracted values, can be only leaves.
The parent-child relation in the output tree is set with optional predicates.
In spite of the fact that OXPath supports XPath's basic types such as boolean, integer and string, all extracted values are transformed into strings.

The structure of the output tree reflects the structure of the extraction markers in the OXPath expression, but not that of the input tree.
Thus, with the use of axes with a direction opposite to the child XPath axis (e.g., parent or ancestor), the resulting tree can considerably differs from the input tree of visited web pages.

\begin{example} 

	Figure~\ref{fig:dblp:outTreeExample} illustrates the output tree produced for the OXPath expression from Example~\ref{exmpl:extractionMarkers} on page \pageref{exmpl:extractionMarkers} (see a wrapper in Listing~\ref{lst:oxpath:outTreeExample} and an XML output in Listing~\ref{lst:xml:outTreeExample}).
	As we can see the root $\top$ contains two child nodes, \texttt{title} (an attribute node) and \texttt{journals} (a record node).
	\texttt{Journals} contains \texttt{journal} as child record nodes each of each contains attribute nodes \texttt{name} and \texttt{url}.
	This example also illustrates how parent-child relation is defined with the use of predicates in OXPath.

\begin{figure}[t]
	\centering
	\includegraphics[width=10cm]{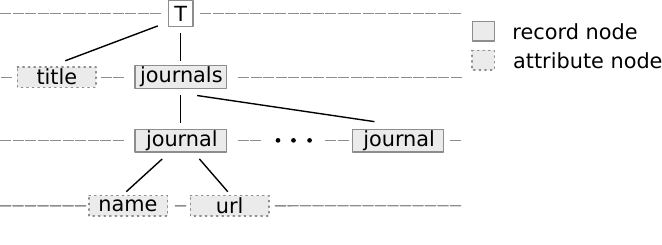}
	\caption{An OXPath output tree for data extracted with wrapper from Listing~\ref{lst:oxpath:outTreeExample}\label{fig:dblp:outTreeExample}}
\end{figure}
	
\begin{lstlisting}[language=OXPath,caption=Example of an OXPath expression for extracting journals from dblp website, label=lst:oxpath:outTreeExample]
  doc("http://dblp.dagstuhl.de/db/journals/?pos=1")
    //header[@class~="headline"]:<title=string(./h1)>
    :<journals>
      [./..//div[@class="hide-body"]//a:<journal>
        [.:<name=string(.)>:<url=string(@href)>]
      ]
\end{lstlisting}

\begin{lstlisting}[language=xml,caption=An XML output produced by converting output tree in Figure~\ref{fig:dblp:outTreeExample} produced by the wrapper in Listing~\ref{lst:oxpath:outTreeExample}, label=lst:xml:outTreeExample]
<results>
  <title>Computer Science Journals</title>
  <journals>
    <journal>
      <name>3-D Information Modeling; International Journal of  ... (IJ3DIM)</name>
      <url>http://dblp.dagstuhl.de/db/journals/ij3dim</url>
    </journal>
    <journal>
      <name>4OR: Quarterly Journal of the Belgian, French and Italian Operations Research Societies</name>
      <url>http://dblp.dagstuhl.de/db/journals/4or</url>
    </journal>
    ...
  <journals>
</results>
\end{lstlisting}
\end{example}


%





\section{OXPath Integration into Java Applications}\label{sec:integration}

OXPath can also be integrated as a Java library into a software project.
To achieve this add the \OXPathCli JAR file into the project's \texttt{CLASSPATH}.



\begin{minipage}{\textwidth}
  \begin{lstlisting}[%
    language=java,
    caption={Minimal working example for the integration of OXPath in Java \label{java:dev:integration:mwe}}]
  package uk.ac.ox.cs.diadem.oxpath.oxpath-example;
  import java.io.BufferedReader; //...

  public class Application {
    public static void main(String[] args) {
      try {    
        // configure browser builder and create browser instance
        WebBrowserBuilder builder = new WebBrowserBuilder();
        RunConfigurationAdapter runcofig = builder.getRunConfiguration();
        runcofig.setEnablePlugins(false);
        runcofig.setDisabledContentTypes(WebBrowser.ContentType.IMAGE);
        runcofig.setXvfbMode(true);
        runcofig.setWidth(1920);
        runcofig.setHeight(1080);
        browser = builder.build();

        final XMLOutputHandler outputHandler = new XMLOutputHandler();
        // retrieve a query from a resource
        final Reader input = new BufferedReader(
            new InputStreamReader(Application.class.
                getResourceAsStream("example.oxp")));
        // invoke OXPath
        OXPath.ENGINE.evaluate(input, browser, outputHandler);
        // prints XML output to "standard" output stream
        System.out.println(outputHandler.asString());
      } catch (IOException e) {
        e.printStackTrace();
      } catch (OXPathRuntimeException e) {
        e.printStackTrace();
      } catch (WebAPIException e) {
        e.printStackTrace();
      } finally {
        // shutdown the browser
        browser.shutdown();
  } } }
  \end{lstlisting}
\end{minipage}

A minimal working example in Listing~\ref{java:dev:integration:mwe} shows which basic building blocks are required by the OXPath API to execute an OXPath expression.
\begin{enumerate}
	\item A \texttt{WebBrowser} instance created with the help of a \texttt{WebBrowserBuilder} (lines~8--15).
	\item Any output handler implementing the \texttt{IAbstractOutputHandler} interface to convert the tree-like output tree into a desired format (line~17).
	\item The OXPath expression needs to be provided as a \texttt{Reader}, \texttt{File}, or \texttt{String} object (lines~19--21).
\end{enumerate}
In line~23, OXPath engine is executed for the specified components and on line 34 the web browser is closed.

Different components of OXPath are united under the umbrella name \OXPathProject.
The current version of \OXPath (\OXPathVersion) and \OXPathCli (\OXPathCliVersion) are generated by \OXPathProjectWithVersion.
It consists of the following main components:
\begin{description}
	\item[OXPath Core (v.\OXPathVersion)] implementing the core functionality of the OXPath language.
	\item[\WebAPI (v.\WebAPIVersion)] implementing an interface to web browsers based on Selenium~2.53.1 (only Firefox~47.0.1 is currently supported).
	\item[Output Handlers] are a set of modules for serialising the output tree of OXPath into different formats.
	The following output handlers are available:
	\texttt{XMLOutputHandler} for XML (see Section~\ref{sec:xmlOutput} on page~\pageref{sec:xmlOutput}),
	\texttt{JsonOutputHandler} for JSON (see Section~\ref{sec:jsonOutput} on page~\pageref{sec:jsonOutput}),
	\texttt{RecStreamCSVOutputHandler} for \texttt{rscsv} (see Section~\ref{sec:csvOutput} on page~\pageref{sec:csvOutput}),
	\texttt{HierarchyCSVOutputHandler} for \texttt{hcsv}, 
	\\ \texttt{RecStreamJDBCOutputHandler} for \texttt{rsjdbc} (see Section~\ref{sec:dbOutput} on page~\pageref{sec:dbOutput}), and
	\texttt{HierarchyJDBCOutputHandler} for \texttt{hjdbc}. 
	\item[\OXPathCli (v.\OXPathCliVersion)] is a command line interface for OXPath.
\end{description}

Java documentation API is available at \OXPathProjectAPI.


\chapter{OXPath in Action}



This chapter describes the process of writing OXPath wrappers by example of the ACL Anthology website from the publication domain.
We consider two main web data extraction scenarios:
1) extraction with a focused crawling through web pages accessible by links from the web page content (Section~\ref{subsec:oia:start:acl:extract}), and
2) extraction with querying a web interface (Sections~\ref{subsec:oia:start:acl:search}, \ref{subsec:oia:start:acl:combine}).

\section{Focused Crawling and Extracting from the ACL Anthology}\label{subsec:oia:start:acl:extract}

\begin{figure}[tbp]
	\centering
	\includegraphics[width=12cm]{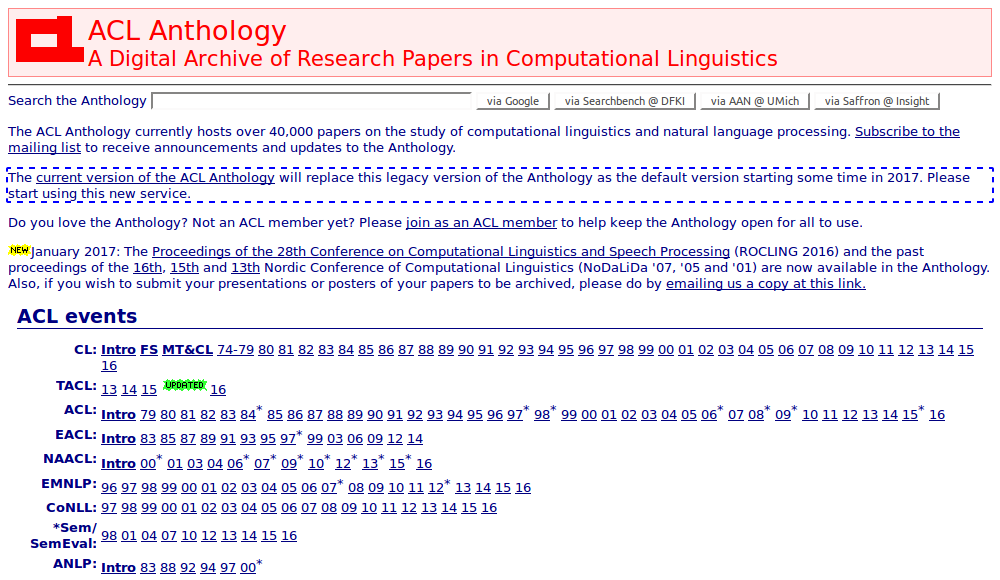}
	\caption{Homepage of the ACL Anthology (\url{http://aclweb.org/anthology/})\label{fig:oia:start:acl:dl}}
\end{figure}

\begin{figure}[tbp]
	\centering
	\includegraphics[width=12cm]{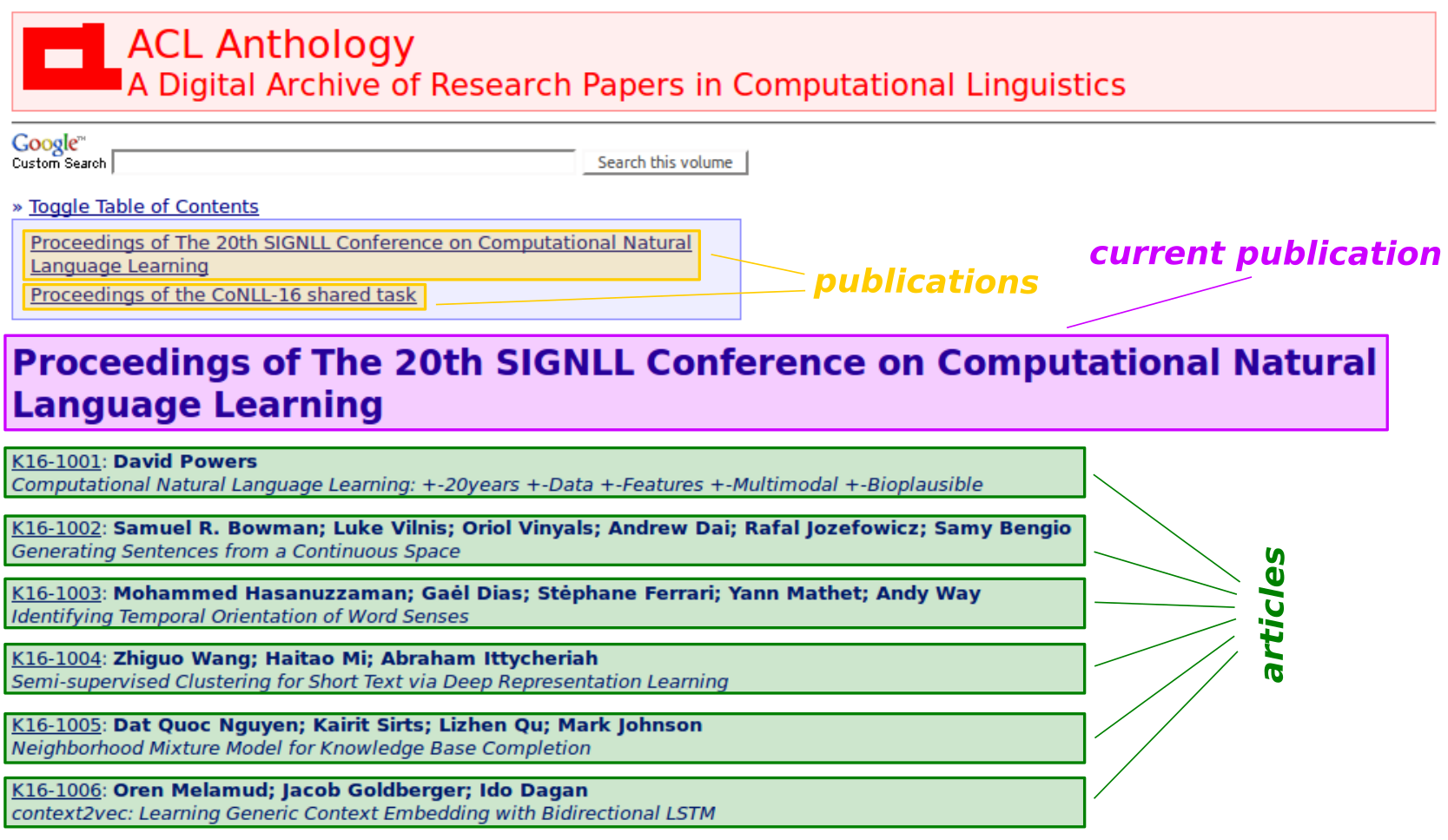}
	\caption{Proceedings of CoNLL 2016\label{fig:oia:start:acl:conll:16}}
\end{figure}

The \emph{Association for Computational Linguistics} (ACL) gives an overview of its events in its digital library, the \emph{ACL Anthology} (see Figure \ref{fig:oia:start:acl:dl}).
The hompage of the ACL Anthology contains a collection of links associated with individual events.
These links lead to publications with very simply structured tables of content, such as in Figure \ref{fig:oia:start:acl:conll:16} for the \emph{Proceedings of the 20th SIGNLL Conference on Computational Natural Language Learning} (CoNLL~2016).
In this section, we consider an approach to crawling ACL events and extracting metadata of publications, presenting an iterative process of writing a wrapper in details.
Furthermore, for the demonstration purposes, we assume that each event page has similarly structured DOM trees. 


Devising of an OXPath expression always starts with analysing the DOM tree of relevant web pages and designing a desired structure of entities or records to be extracted.
In particular, in Listing~\ref{html:oia:start:acl:conll:16}, illustrating an HTML snippet of the CoNLL~2016 web page's DOM tree (see Figure~\ref{fig:oia:start:acl:conll:16}), an {\color{MaterialOrange}\texttt{h1}} element with a title of a publication is followed by sibling elements {\color{MaterialOrange}\texttt{p}} representing a list of articles.
Moreover, all these {\color{MaterialOrange}\texttt{h1}} and {\color{MaterialOrange}\texttt{p}} elements are siblings.
Taking into account such peculiarity of the DOM tree structure and limitations of XPath~1.0, the expected output from OXPath in a form of XML can be as in Listing~\ref{xml:oia:start:acl:conll:16:0}.

\begin{minipage}{\textwidth}
	\newcommand*{\CommonPath}{users/chapters/listings/oxpathInAction/html}
\begin{lstlisting}[%
	language=html,
	linerange=1-30,
	caption={A snippet of a serialised HTML code of the CoNLL~2016 event page's DOM tree\label{html:oia:start:acl:conll:16}}
	]
<html><!--[...]-->
  <div id="content"><!--[...]-->
    <h1>Proceedings of The 20th SIGNLL Conference on Computational Natural Language Learning</h1><!--[...]-->
    <p>
      <a href="K16-1001.pdf">K16-1004</a>
      :
      <b>      <first>Zhiguo</first>      <last>Wang</last>;       <first>Haitao </first>      <last>Mi</last>;       <!--[...]--></b>
      <br>
      <i>Semi-supervised Clustering for Short Text via Deep Representation Learning</i>
    </p><!--[...]-->
    <h1><!--[title of a publication]--></h1><!--[...]-->
    <p><!--[article]--></p><!--[...]-->
    </div>
<!--[...]--><html>
\end{lstlisting}
\end{minipage}

\begin{minipage}{\textwidth}
	\newcommand*{\CommonPath}{users/chapters/listings/oxpathInAction}
\begin{lstlisting}[%
	language=xml,
	caption={A structure of publications extracted from CoNLL 2016\label{xml:oia:start:acl:conll:16:0}}
	]
<results>
  <event>
    <article>
      <title></title>
      <authors></authors>
      <pdf></pdf>
      <publication></publication>
    </article><!--[...]-->
  </event><!--[...]-->
</results>
\end{lstlisting}
\end{minipage}

As we can see, it has a list of {\color{MaterialGreen}\texttt{event}}s.
Each event node contains {\color{MaterialGreen}\texttt{article}}s from an event page comprising the title of the corresponding {\color{MaterialGreen}\texttt{publication}}, a {\color{MaterialGreen}\texttt{title}}, {\color{MaterialGreen}\texttt{authors}}, and a URL to a {\color{MaterialGreen}\texttt{pdf}} file.

For most web data extraction scenarios we first need to identify a main entity or record to be extracted (i.e., an {\color{MaterialGreen}\texttt{article}}) and then define the focused crawling or navigation strategy.

\subsection{Extracting from the ACL Anthology}


As we can see, all articles, wrapped by the {\color{MaterialOrange}\texttt{p}} element, are in the unique {\color{MaterialOrange}\texttt{div}} as shown in line 2 of Listing~\ref{html:oia:start:acl:conll:16}.
An interim OXPath expression matching relevant {\color{MaterialOrange}\texttt{p}} elements and outputing empty {\color{MaterialGreen}\texttt{article}} element nodes will look as presented in Listing~\ref{oxp:oia:start:acl:conll:16:1}.


\begin{minipage}{\textwidth}
  \newcommand*{\CommonPath}{users/chapters/listings/oxpathInAction}
\begin{lstlisting}[%
    language=oxpath,
    caption={Basic article identification with OXPath for the proceedings of CoNLL 2016 and respective XML output\label{oxp:oia:start:acl:conll:16:1}}
  ]
doc('https://www.aclweb.org/anthology/K/K16/')
  //div[@id='content']/p:<article>
\end{lstlisting}
\begin{lstlisting}[%
    language=xml,
  ]
<results>
	<article/>
	<article/>
	<article/><!--[...]-->
</results>
\end{lstlisting}
\end{minipage}

Inside each {\color{MaterialOrange}\texttt{p}} element (see Listing~\ref{html:oia:start:acl:conll:16}, lines~4--10), several data fields are available in separate subordinate elements.
In particular, each {\color{MaterialOrange}\texttt{a}} element provides a link to the PDF file of the publication (line~5), the following {\color{MaterialOrange}\texttt{b}} element lists the authors (line~7), and the {\color{MaterialOrange}\texttt{i}} element contains the publication title (line~9).
In order to incorporate these data fields into our result tree as children of the {\color{MaterialGreen}\texttt{article}}-nodes, predicate statements should be used.
Each predicate statement in this case contains a relative XPath expression followed by an extraction marker as in Listing~\ref{oxp:oia:start:acl:conll:16:2} (lines 3--5).

The table of content also attributes each article to a specific publication.
The latter can be extracted for each article with the help of XPath expression {\color{MaterialOrange}\texttt{./preceding::h1[1]}} (see line~6), in which the axis {\color{MaterialOrange}\texttt{preceding}} selects {\color{MaterialOrange}\texttt{h1}} nodes in a reverse document order and the position predicate {\color{MaterialOrange}\texttt{[1]}} limits the selection to the first preceeding element from the result set.


\begin{minipage}{\textwidth}
  \newcommand*{\CommonPath}{users/chapters/listings/oxpathInAction}
\begin{lstlisting}[%
    language=oxpath,
    caption={Article extraction with OXPath for the proceedings of CoNLL 2016 and respective XML output. In this example, \textvisiblespace{} is used to indicate multiple whitespaces\label{oxp:oia:start:acl:conll:16:2}}
  ]
doc('https://www.aclweb.org/anthology/K/K16/')
  //div[@id='content']/p:<article>
    [./i:<title=string(.)>]
    [./b:<authors=string(.)>]
    [./a:<pdf=string(@href)>]
    [./preceding::h1[1]:<publication=string(.)>]
\end{lstlisting}
\begin{lstlisting}[%
    language=xml, escapechar=ä
  ]
<results>
  <!--[...]-->
  <article>
    <title>Semi-supervised Clustering for Short Text via Deep Representation Learning</title>
    <authors>ä\textvisiblespace\textvisiblespace\textvisiblespace\textvisiblespace\textvisiblespace\textvisiblespaceäZhiguoä\textvisiblespace\textvisiblespace\textvisiblespace\textvisiblespace\textvisiblespace\textvisiblespaceäWang;ä\textvisiblespace\textvisiblespace\textvisiblespace\textvisiblespace\textvisiblespace\textvisiblespaceäHaitaoä\textvisiblespace\textvisiblespace\textvisiblespace\textvisiblespace\textvisiblespace\textvisiblespaceäMi;ä\textvisiblespace\textvisiblespace\textvisiblespace\textvisiblespace\textvisiblespace\textvisiblespaceä[...]</authors>
    <pdf>K16-1001.pdf</pdf>
    <publication>Proceedings of The 20th SIGNLL Conference on Computational Natural Language Learning</publication>
  </article><!--[...]-->
</results>
\end{lstlisting}
\end{minipage}

A closer look at the XML output in Listing \ref{oxp:oia:start:acl:conll:16:2} raises the question of refinement options:
(1)~the value of the {\color{MaterialDeepOrange}\texttt{href}} attribute of the {\color{MaterialOrange}\texttt{a}} element contains a relative URL path to the PDF file of an article, and it should be respectively replaced by the absolute URL path;
(2)~unnecessary whitespaces should be eliminated from the list of authors.

These issues can be addressed with the help of (O)XPath functions.
In particular, (1) {\color{MaterialGreen}\texttt{qualify-url()}} completes the partial or relative URL with respect to the URL of a web page (see Listing \ref{oxp:oia:start:acl:conll:16:4}, line 5);
and (2) {\color{MaterialGreen}\texttt{normalize-space()}} removes excessive whitespace (see line 4).


\begin{minipage}{\textwidth}
  \newcommand*{\CommonPath}{users/chapters/listings/oxpathInAction}
\begin{lstlisting}[%
    language=oxpath,
    caption={Refined article extraction with OXPath for the proceedings of CoNLL 2016 and respective XML output\label{oxp:oia:start:acl:conll:16:4}}
  ]
doc('https://www.aclweb.org/anthology/K/K16/')
  //div[@id='content']/p:<article>
    [./i:<title=string(.)>]
    [./b:<authors=normalize-space(.)>]
    [./a:<pdf=qualify-url(@href)>]
    [./preceding::h1[1]:<publication=string(.)>]
\end{lstlisting}
\begin{lstlisting}[%
    language=xml,
  ]
<results>
  <!--[...]-->
  <article>
  	<title>Semi-supervised Clustering for Short Text via Deep Representation Learning</title>
  	<authors>Zhiguo Wang; Haitao Mi; [...]</authors>
    <pdf>https://www.aclweb.org/anthology/K/K16/K16-1001.pdf</pdf>
    <publication>Proceedings of The 20th SIGNLL Conference on Computational Natural Language Learning</publication>
  </article><!--[...]-->
</results>
\end{lstlisting}
\end{minipage}

\subsection{Extracting with Focused Crawling}

Our OXPath expression can be extended with a focused crawling functionality to navigate all ACL events and extract their publications.
To achieve that, we need to add a multi-way navigation into the wrapper.

As we can see in Listing~\ref{oxp:oia:start:acl:conll:16:5-nav}, the home page of the ACL Antology is chosen as the seed web page (see line~1).
In line~2, we
1)~select relevant event elements with an XPath expression {\color{MaterialOrange}\texttt{//table//tr/td/a}},
2)~navigate to a web page with corresponding articles using an absolute action step {\color{MaterialRed}\texttt{\{click/\}}},
and 3)~extract articles (lines~3--8), nested in the {\color{MaterialGreen}\texttt{event}} node.

\begin{minipage}{\textwidth}
	\newcommand*{\CommonPath}{users/chapters/listings/oxpathInAction}
\begin{lstlisting}[%
	language=oxpath,
	caption={Article extraction with OXPath for the ACL proceedings and respective XML output\label{oxp:oia:start:acl:conll:16:5-nav}}
	]
doc('http://aclweb.org/anthology/')
  //table//tr/td/a/{click/}/.:<event>
    [.//div[@id='content']/p:<article>
        [./i:<title=string(.)>]
        [./b:<authors=normalize-space(.)>]
        [./a:<pdf=qualify-url(@href)>]
        [./preceding::h1[1]:<publication=string(.)>]
    ]
\end{lstlisting}
\begin{lstlisting}[%
	language=xml,
	]
<results>
  <!--[...]-->
  <event>
    <article>
      <title>Semi-supervised Clustering for Short Text via Deep Representation Learning</title>
  	  <authors>Zhiguo Wang; Haitao Mi; [...]</authors>
      <pdf>https://www.aclweb.org/anthology/K/K16/K16-1001.pdf</pdf>
      <publication>Proceedings of The 20th SIGNLL Conference on Computational Natural Language Learning</publication>
    </article><!--[...]-->
  </event><!--[...]-->
</results>
\end{lstlisting}
\end{minipage}

\section{Querying the ACL Anthology}\label{subsec:oia:start:acl:search}

In contrast to the focused web crawling on the surface of the Web, some web data extraction scenarios can require interaction with web interfaces, translating web form requests into queries for back-end databases.

\begin{figure}
	\centering
	\includegraphics[width=10cm]{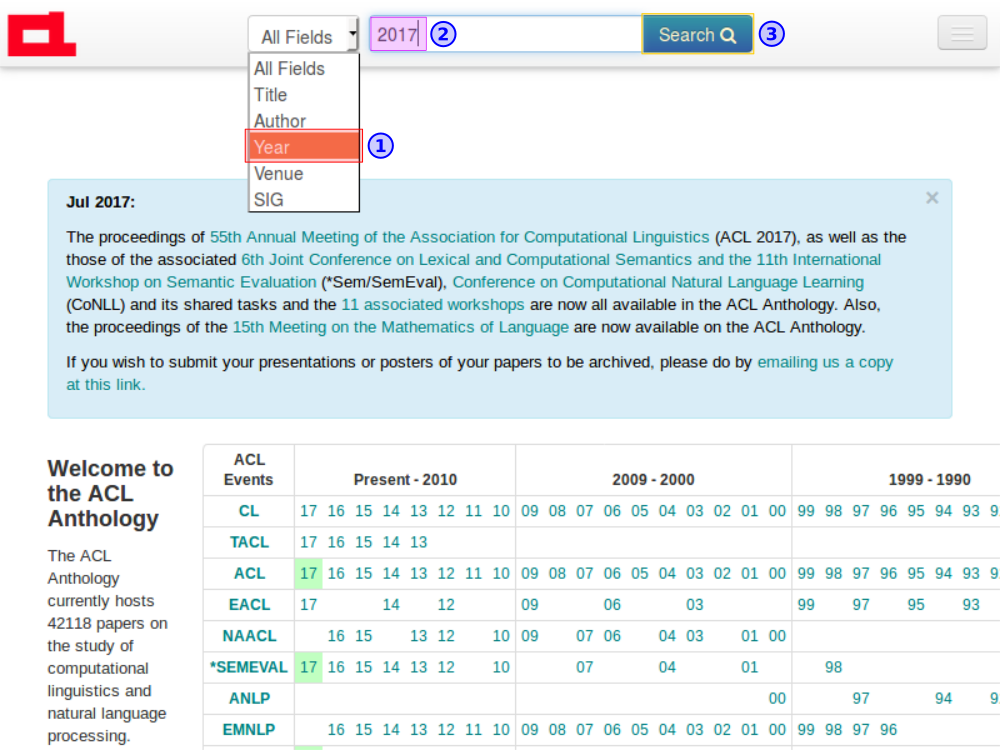}
	\caption{Search with the web form of the ACL Anthology\label{fig:oia:start:acl:info:0}}
\end{figure}

\begin{figure}
	\centering
	\includegraphics[width=10cm]{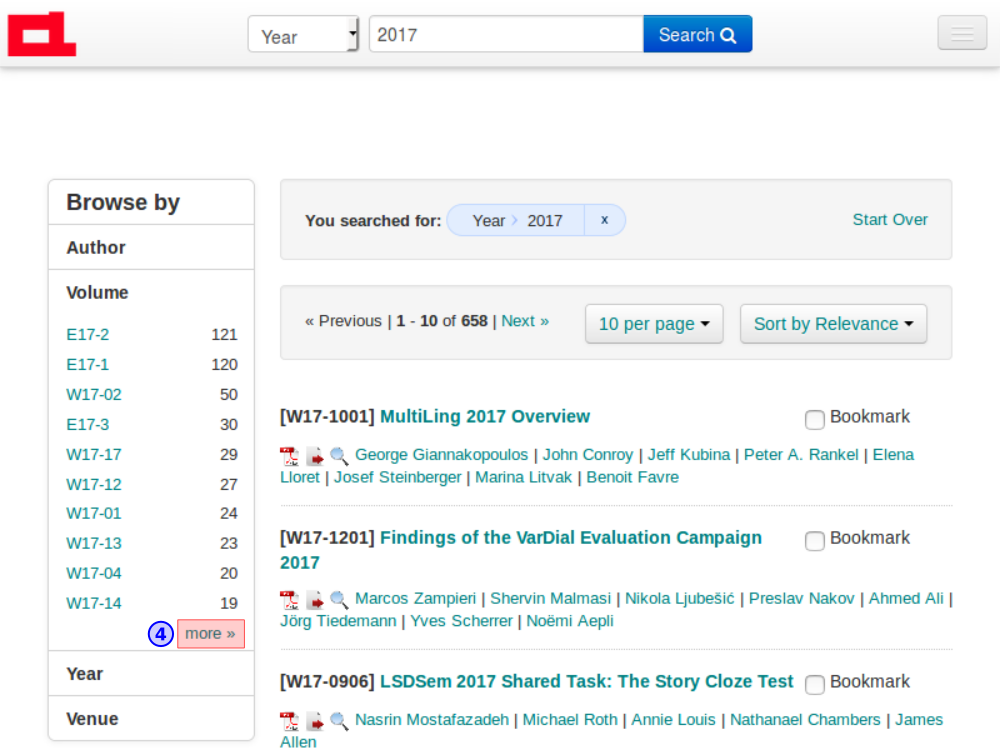}
	\caption{Search results view of the ACL Anthology\label{fig:oia:start:acl:info:2}}
\end{figure}

\begin{figure}
	\centering
	\includegraphics[width=10cm]{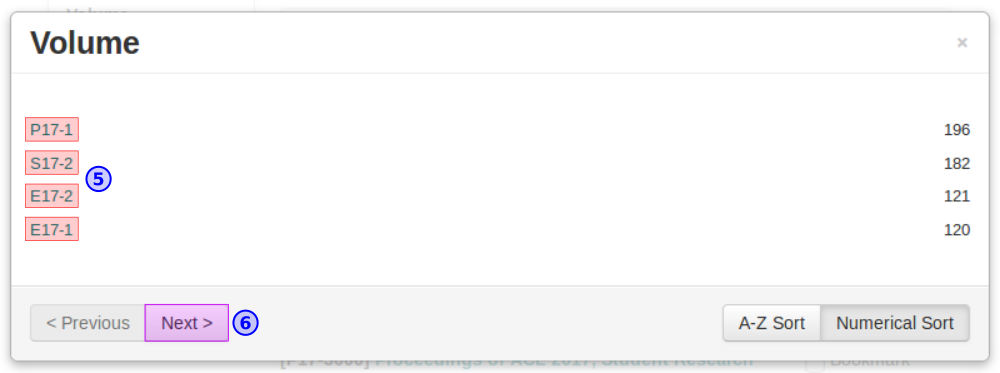}
	\caption{List of all volumes published in 2017 in the ACL Anthology\label{fig:oia:start:acl:info:3}}
\end{figure}

With its newer interface, the ACL Anthology also offers a search web form.
The following scenario gives an example of searching all ACL volumes published in 2017:
(i) we send a search query to the ACL Anthology via its interface to get all volumes published in 2017 (see Figure~\ref{fig:oia:start:acl:info:0});
(ii) on the result page we click the link \enquote{more} within the browse facet \enquote{volume} (see Figure~\ref{fig:oia:start:acl:info:2}), which renders a modal window as in Figure~\ref{fig:oia:start:acl:info:3} with a complete list of volumes;
(iii) we iteratively extract all available anthology volumes by clicking the next button.

An OXPath expression in Listing \ref{oxp:oia:start:acl:info:0} (page \pageref{oxp:oia:start:acl:info:0}) implements this scenario.

\noindent \textbf{(i)}
%
\begin{enumerate}
	\item The filter is set into the year-based search mode (line 2).
	\item The term \enquote{2017} is entered into the search field (line 3).
	\item A search is triggered by clicking the search button of the web form (line 4).\\
	Instead of clicking the search button we can also simulate pressing the enter button on the search field.
	This can be achieved with the absolute action {\color{MaterialRed}\texttt{\{pressenter/\}}} and replacing the absolute type action with a contextual action \lstinline[language=oxpath]|{"2017"}|.
\end{enumerate}
%


\begin{minipage}{\textwidth}
  \newcommand*{\CommonPath}{users/chapters/listings/oxpathInAction}
\begin{lstlisting}[%
    language=oxpath,
    caption={An OXPath expression for searching all volumes published in 2017 in the ACL Anthology and its XML output\label{oxp:oia:start:acl:info:0}}
  ]
doc('http://aclanthology.info')
  //*[@id='search_field']/option[@value='publish_date']/{click /}
  //*[@id='q']/{"2017" /}
  //*[@id='search']/{click /}
  //*[@id='facets']//*[contains(@class, 'volume_anthology')]
      //a[contains(@class, 'more')]/{click /}
    /?!(!?//*[@id='ajax-modal']//a[@rel='next']/{click /}?!)*!?
      //*[@id='ajax-modal']//a[@class='facet_select']:<volume>
        [.:<code=normalize-space(.)>]
        [.:<url=qualify-url(@href)>]
\end{lstlisting}
\begin{lstlisting}[%
    language=xml,
  ]
<?xml version="1.1" encoding="UTF-8"?>
<results>
    <volume>
        <code>E17-2</code>
        <url>http://aclanthology.info/catalog?...</url>
    </volume>
    <volume>
        <code>E17-1</code>
        <url>http://aclanthology.info/catalog?...</url>
    </volume>
    <!--[...]-->
</results>
\end{lstlisting}
\end{minipage}

\noindent \textbf{(ii)}
\begin{enumerate}
	\setcounter{enumi}{3}
	\item In lines 5 and 6, we initiate a click action for the link \enquote{more}. This action triggers AJAX request, which returns a list of all relevant volumes, rendered further in a modal window.\\
	Initially the browse filter volume is folded, and the OXPath hence cannot simulate user interaction by simply clicking the element at specific location within the viewport.
	Instead, in this case OXPath fires all relevant mouse DOM events directly on the DOM tree triggering corresponding browser actions.
	Reducing the number of interaction we increase the robustness of the wrapper.
\end{enumerate}

\noindent \textbf{(iii)}
\begin{enumerate}
	\setcounter{enumi}{4}
	\item With the Kleene star (line 7), the expression iterates over result pages in the modal window and extracts anthology volume records with their corresponding code and URL (lines~8--10).
\end{enumerate}


An XML result of the evaluation is presented below the OXPath wrapper in Listing \ref{oxp:oia:start:acl:info:0}.
Listings \ref{html:oia:start:acl:info:0}, \ref{html:oia:start:acl:info:2}, and \ref{html:oia:start:acl:info:3} provide excerpts from the HTML sources of the relevant pages of the ACL Anthology to illustrate the elements selected by the OXPath expression for steps (i), (ii), and (iii), respectively.

\begin{minipage}{\textwidth}
  \newcommand*{\CommonPath}{users/chapters/listings/oxpathInAction/html}
\begin{lstlisting}[%
    language=html,
    linerange=1-30,
    caption={HTML source for the homepage of the ACL Anthology\label{html:oia:start:acl:info:0}}
  ]
<html><!--[...]-->
  <select id="search_field" class="..."><!--[...]-->
    <option value="publish_date">Year</option>
  <!--[...]--></select>
  <input id="q" class="..."/>
  <button id="search" class="..." type="submit">
  <!--[...]--></button>
<!--[...]--><html>
\end{lstlisting}
\end{minipage}

\begin{minipage}{\textwidth}
  \newcommand*{\CommonPath}{users/chapters/listings/oxpathInAction/html}
\begin{lstlisting}[%
    language=html,
    linerange=1-30,
    caption={HTML source for the search result view of the ACL Anthology\label{html:oia:start:acl:info:2}}
  ]
<html><!--[...]-->
  <div id="facets" class="facets sidenav"><!--[...]-->
  <div class="facet_limit blacklight-volume_anthology"><!--[...]-->
    <a class="more_facets_link" href="..."><!--[...]--></a>
  <!--[...]--></div>
<!--[...]--><html>
\end{lstlisting}
\end{minipage}

\begin{minipage}{\textwidth}
  \newcommand*{\CommonPath}{users/chapters/listings/oxpathInAction/html}
\begin{lstlisting}[%
    language=html,
    linerange=1-30,
    caption={HTML source for the filter modal of the ACL Anthology\label{html:oia:start:acl:info:3}}
  ]
<html><!--[...]-->
  <div id="ajax-modal" class="..."><!--[...]-->
    <ul class="facet_extended_list"><!--[...]-->
      <li>
        <a class="facet_select" href="...">P17-1</a>
        <!--[...]--></li>
    <!--[...]--></ul>
  <!--[...]--></div>
<!--[...]--><html>
\end{lstlisting}
\end{minipage}

\section{Shortcut to Search Facets}\label{subsec:oia:start:acl:shortcut}

\begin{figure}[!h]
	\centering
	\includegraphics[width=12cm]{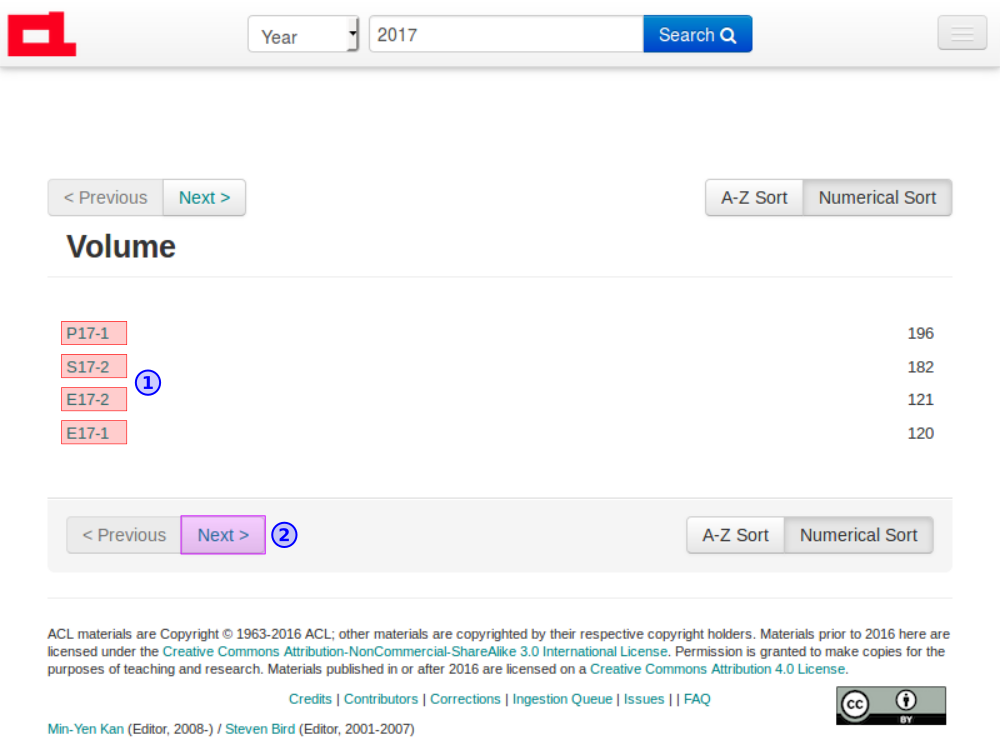}
	\caption{Advanced search volume facet of the ACL Anthology\label{fig:oia:start:acl:info:3b}}
\end{figure}

The routing of a website may provide shortcuts to its individual facets. 
This can considerably simplify the wrapper.

For example, for the ACL Anthology the URL leading us to the search results as in Figure~\ref{fig:oia:start:acl:info:3} (page~\pageref{fig:oia:start:acl:info:3}) is exposed by the link \enquote{more} in Figure~\ref{fig:oia:start:acl:info:2}.
Entering this link in the web browser will render a web page as in Figure~\ref{fig:oia:start:acl:info:3b}.
A set of possible URL query parameters is also included into this URL, e.g. \texttt{facet.sort=count}, \texttt{q=2017}, \texttt{search\_field=publish\_date}, etc.
A wrapper for extracting data from a web page with this direct link is presented in Listing~\ref{oxp:oia:start:acl:info:1} and an HTML snippet of the rendered page is in Listing~\ref{html:oia:start:acl:info:3b}.

\begin{minipage}{\textwidth}
  \newcommand*{\CommonPath}{users/chapters/listings/oxpathInAction}
\begin{lstlisting}[%
    language=oxpath,
    caption={OXPath expression for extracting the list of volumes of a given year from the volume search facet of the ACL Anthology \label{oxp:oia:start:acl:info:1}}
  ]
doc('http://.../catalog/facet/volume_anthology?q=2017&...')
  /?!(!?//*[@id='content']//*[@class='facet_pagination top']//a[@rel='next']/{click[wait=2] /}?!)*!?
    //*[@id='content']//a[@class='facet_select']:<volume>
      [.:<code=normalize-space(.)>]
      [.:<url=qualify-url(@href)>]
\end{lstlisting}
\end{minipage}

\begin{minipage}{\textwidth}
  \newcommand*{\CommonPath}{users/chapters/listings/oxpathInAction}
\begin{lstlisting}[%
    language=html,
    linerange=1-30,
    caption={HTML source for the volume search facet of the ACL Anthology\label{html:oia:start:acl:info:3b}}
  ]
<html><!--[...]-->
  <div id="content" class="row"><!--[...]-->
    <div class="facet_pagination top"><!--[...]-->
        <a class="btn" href="..." rel="next">Next ></a>
    <!--[...]--></div>
    <ul class="facet_extended_list"><!--[...]-->
      <li>
        <a class="facet_select" href="...">P17-1</a>
        <!--[...]--></li>
    <!--[...]--></ul>
  <!--[...]--></div>
<!--[...]--><html>
\end{lstlisting}
\end{minipage}



\section{Querying and Extracting from the ACL Anthology}\label{subsec:oia:start:acl:combine}

In Section~\ref{subsec:oia:start:acl:extract}, we considered an example of extracting all articles from the ACL Anthology with the use of a focused crawling scenario.
In contrast, in this section we specifically extract articles published in 2017 by querying a search form of the ACL Anthology.

\begin{figure}[h]
	\centering
	\includegraphics[width=7.8cm]{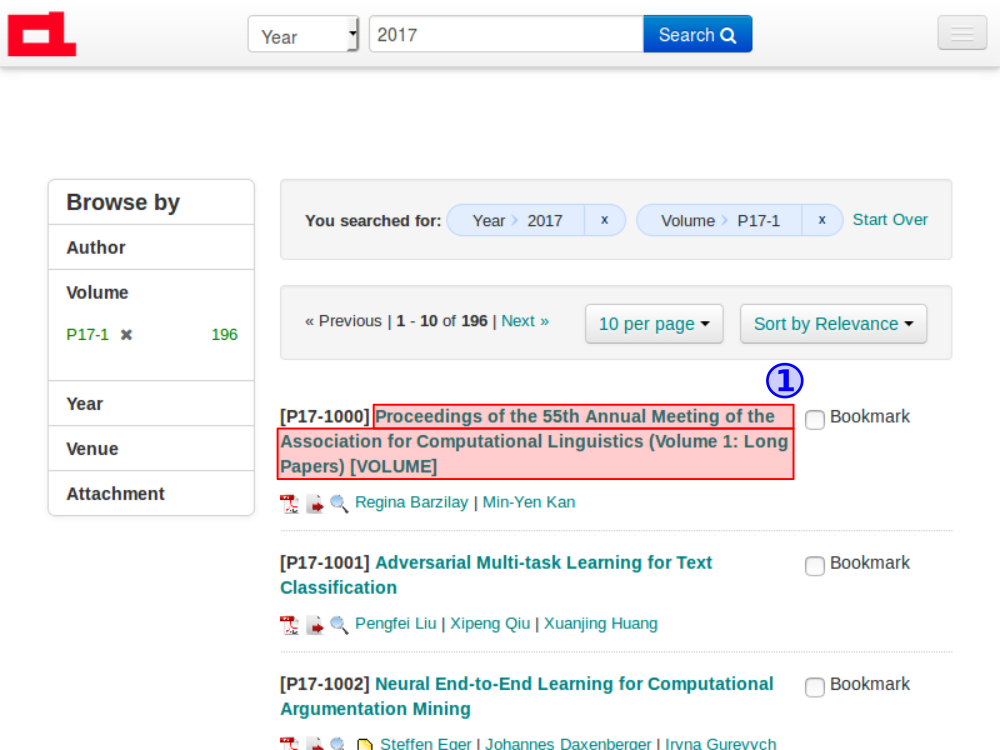}
	\caption{Result view for an individual volume in the ACL Anthology \label{fig:oia:start:acl:info:4b}}
\end{figure}

\begin{figure}[h]
	\centering
	\begin{minipage}[t]{0.49\textwidth}
		\centering
		\includegraphics[width=7.8cm]{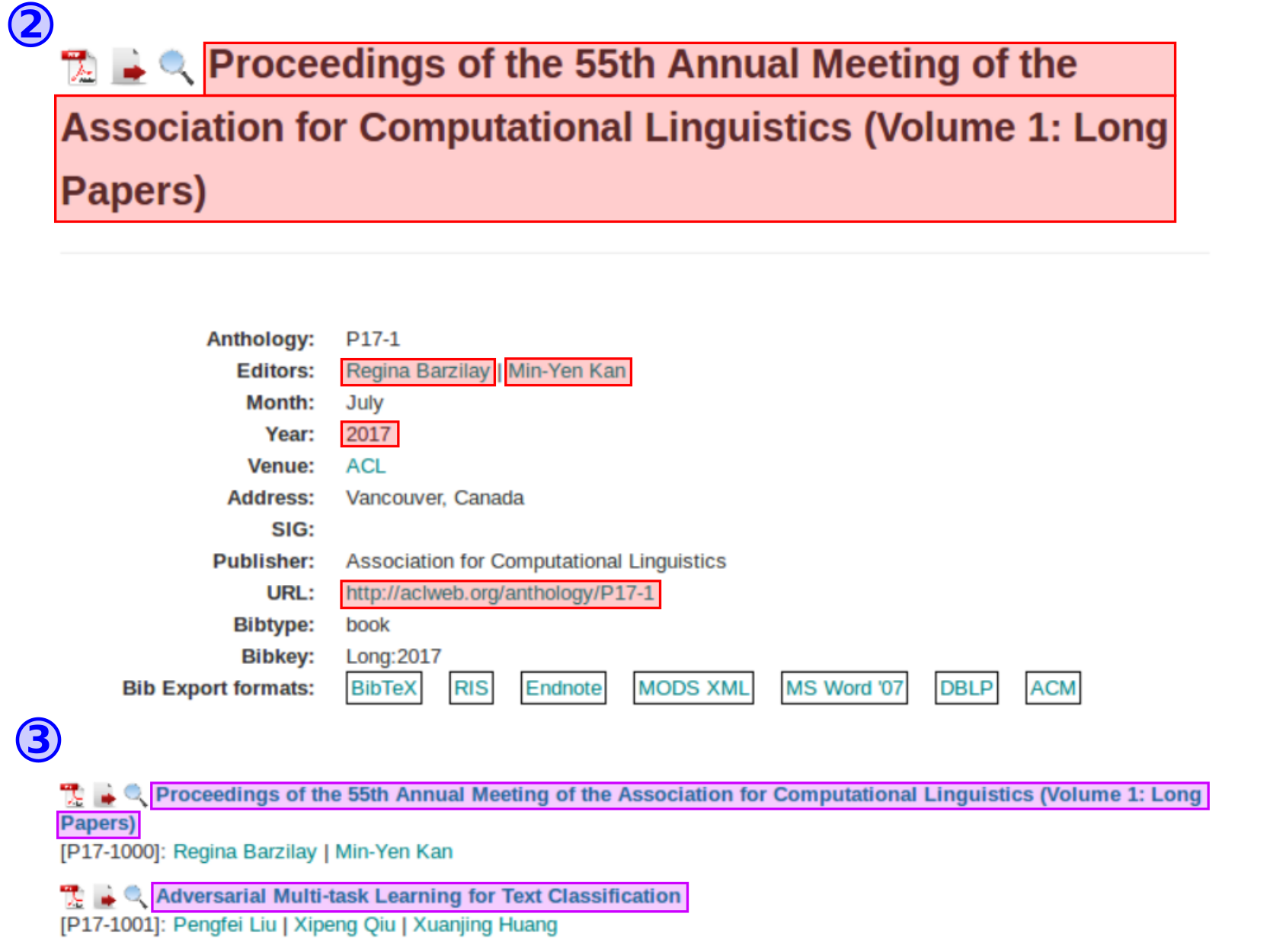}
		\caption{Table of contents for a volume in the ACL Anthology\label{fig:oia:start:acl:info:5}}
	\end{minipage}
	\hfill
	\begin{minipage}[t]{0.49\textwidth}
		\centering
		\includegraphics[width=7.8cm]{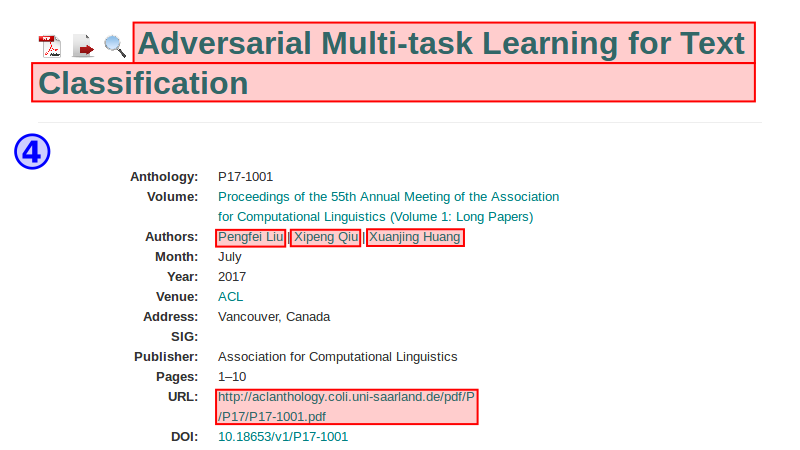}
		\caption{Detail page of an individual article in the ACL Anthology; visited for extraction \label{fig:oia:start:acl:info:6b}}
	\end{minipage}
\end{figure}

For querying we use an example from Section~\ref{subsec:oia:start:acl:shortcut}, which navigates over anthology volumes from 2017, obtained directly by the URL.
Each anthology volume is a link to a web page with corresponding publication entries as illustrated in Figure~\ref{fig:oia:start:acl:info:4b}.
The first item on this page represents all conference proceedings and the rest are articles from these proceedings.
All articles from such page can be extracted with the following scenario:
(i) we navigate to a detail page of the conference proceedings, which also contains all articles in one page (without a paginator);
(ii) extracts detailed information of the proceedings and navigate to detail pages of articles (see Figure~\ref{fig:oia:start:acl:info:5});
(iii) Within the article detail page, we extract desired metadata (see Figure~\ref{fig:oia:start:acl:info:6b}).

This scenario is implemented in an OXPath wrapper presented in Listing~\ref{oxp:oia:acl:querying:articles}.
%

\noindent \textbf{(i)}
\begin{enumerate}
	\item We select the first item with conference proceedings and navigate to its detail page (lines 2, 3).
\end{enumerate}

\noindent \textbf{(ii)}
\begin{enumerate}
	\setcounter{enumi}{1}
	\item We extract metadata of the current proceedings, such as the name (line 4), editors (lines 5, 6), year (lines 7, 8), and the URL to the PDF file (lines 9, 10).
	\item For each article selected with a plain XPath selector (line 11), we initiate a click action to navigate to its detail page (line 12).
	With a set of articles returned by the XPath expression and the click action, OXPath perform a multi-way navigation.
	This means that OXPath will navigate back and forward from an opened article detail page back to the proceedings page.
\end{enumerate}

\noindent \textbf{(iii)}
\begin{enumerate}
	\setcounter{enumi}{3}
	\item On each article page we extract required metadata, such as the URL of a page (line 14), title (line 15), authors (lines 16, 17), and URL of the PDF file (lines 18, 19).
\end{enumerate}


\begin{minipage}{\textwidth}
  \newcommand*{\CommonPath}{users/chapters/listings/oxpathInAction}
\begin{lstlisting}[%
    language=oxpath,
    caption={OXPath expression for extracting metadata of a given volume from the new layout of the ACL Anthology \label{oxp:oia:acl:querying:articles}}
  ]
doc('https://.../catalog?q=2017&...')
  //div[@class~='document']//h5[@class~='index_title']/a[@href#='/volumes/']
    /{click/}/.:<publication>
  [.//div[@class~='page-header']/.:<name=normalize-space(.)>]
  [? .//dt[contains(., 'Editors')]/following-sibling::dd[1]
          :<editors=normalize-space(string-join(./a[contains(@href, 'people')], '; '))>]
    [? .//dt[contains(., 'Year')]/following-sibling::dd[1]
          :<year=normalize-space(.)>]
    [? .//dt[contains(., 'URL')]/following-sibling::dd[1]
          :<pdf=normalize-space(.)>]
    //p//strong/a[contains(@href,'papers')]
      [./{click/}
        /.:<article>
          [.:<url=current-url()>]
          [.//div[@class~='page-header']/.:<title=normalize-space(.)>]
          [? .//dt[contains(., 'Authors')]/following-sibling::dd[1]
            :<authors=normalize-space(string-join(./a[contains(@href, 'people')], '; '))>]
          [? .//dt[contains(., 'URL')]/following-sibling::dd[1]
            :<pdf=normalize-space(.)>]
      ]
\end{lstlisting}
\end{minipage}


A complete OXPath expression is presented in Listing~\ref{oxp:oia:acl:querying:all:articles2017} and a snippet of an XML output is in Listing~\ref{xml:oia:acl:querying:all:articles2017}.

\begin{minipage}{\textwidth}
  \newcommand*{\CommonPath}{users/chapters/listings/oxpathInAction}
\begin{lstlisting}[%
    language=oxpath,
    caption={OXPath expression for extracting articles published in 2017 by querying a web form of the ACL Anthology\label{oxp:oia:acl:querying:all:articles2017}}
  ]
doc('http://aclanthology.info')
  //*[@id='search_field']/option[@value='publish_date']/{click /}
  //*[@id='q']/{"2017" /}
  //*[@id='search']/{click /}
  //*[@id='facets']//*[contains(@class, 'volume_anthology')]
      //a[contains(@class, 'more')]/{click /}
    /?!(!?//*[@id='ajax-modal']//a[@rel='next']/{click /}?!)*!?
      //*[@id='ajax-modal']//a[@class='facet_select']:<volume>
        [.:<code=normalize-space(.)>]
        [.:<url=qualify-url(@href)>]
      /{click/}
  //div[@class~='document']//h5[@class~='index_title']/a[@href#='/volumes/']
    /{click/}/.:<publication>
  [.//div[@class~='page-header']/.:<name=normalize-space(.)>]
  [? .//dt[contains(., 'Editors')]/following-sibling::dd[1]
          :<editors=normalize-space(string-join(./a[contains(@href, 'people')], '; '))>]
    [? .//dt[contains(., 'Year')]/following-sibling::dd[1]
          :<year=normalize-space(.)>]
    [? .//dt[contains(., 'URL')]/following-sibling::dd[1]
          :<pdf=normalize-space(.)>]
    //p//strong/a[contains(@href,'papers')]
      [./{click/}
        /.:<article>
          [.:<url=current-url()>]
          [.//div[@class~='page-header']/.:<title=normalize-space(.)>]
          [? .//dt[contains(., 'Authors')]/following-sibling::dd[1]
            :<authors=normalize-space(string-join(./a[contains(@href, 'people')], '; '))>]
          [? .//dt[contains(., 'URL')]/following-sibling::dd[1]
            :<pdf=normalize-space(.)>]
      ]
\end{lstlisting}
\begin{lstlisting}[%
    language=xml,
    caption={An XML output resulting from the OXPath expression from Listing~\ref{oxp:oia:acl:querying:all:articles2017} \label{xml:oia:acl:querying:all:articles2017}}
  ]
<results><!--[...]-->
    <!--[...]--><volume>
        <code>P17-1</code>
        <url>https://aclanthology.coli.uni-saarland.de/catalog?...</url>
        <!--[...]--><publication>
            <name>Proceedings of the 55th Annual Meeting of the Association for Computational Linguistics (Volume 1: Long Papers)</name>
            <editors>Regina Barzilay; Min-Yen Kan</editors>
            <year>2017</year>
            <pdf>http://aclweb.org/anthology/P17-1</pdf>
            <!--[...]--><article>
              <url>http://aclanthology.coli.uni-saarland.de/...</url>
              <title>Adversarial Multi-task Learning for Text Classification</title>
              <authors>Pengfei Liu; Xipeng Qiu; Xuanjing Huang</authors>
              <pdf>http://www.aclweb.org/anthology/P17-1001</pdf>
            <!--[...]--></article>
        <!--[...]--></publication>
    <!--[...]--></volume>
<!--[...]--></results>
\end{lstlisting}
\end{minipage}

\chapter{Conclusion}

In this tutorial, we have introduced the OXPath language for web data extraction. We have explained the main features of the OXPath language by giving a brief overview of XPath -- the language which OXPath is based on -- and introducing the extensions to it that OXPath provides. An exhaustive list of prototypical examples was provided to give some guidelines on how to use OXPath in typical use cases. In addition, the setup of a suitable working environment has been outlined as well as a short introduction into the OXPath API for developers.


\chapter*{Bibliography}

\defbibfilter{collect:articles}{
  type=article or
  type=inproceedings
}

\addcontentsline{toc}{chapter}{\textcolor{Primary1}{Bibliography}}
\phantomsection
\printbibliography[heading=bibempty]



\cleardoublepage
\phantomsection
\setlength{\columnsep}{0.75cm}
\addcontentsline{toc}{chapter}{\textcolor{Primary1}{Index}}
\printindex


\end{document}